%% file: h4542.tex
\documentclass{aa}
\usepackage{txfonts}
\usepackage{natbib}
\usepackage{epsfig}

\bibpunct{(}{)}{;}{a}{}{,}
\usepackage{graphicx}

\newcommand{\teff}{$T_{\rm eff}$}

\newcommand{\logg}{log\,$g$}
\newcommand{\kms}{km\,s$^{-1}$}

\newcommand{\msun}{$M_{\odot}$}
\begin{document}
   \title{Studying the populations of our Galaxy using the kinematics of sdB stars\thanks{Based on data obtained
at the 1.54 m telescope of the European Southern Observatory, runs 58.D-0478, 60.D-0306, 62.L-0220, 64.L-0077,
64.H-0023(A), 65.H-0250, 65.H-0341(A) and 68.D-0192(A)}}

   \subtitle{}

   \author{M. Altmann
          \inst{1,2}
          \and
          H. Edelmann\inst{2}
          \and 
          K.S. de Boer\inst{1}
          }

     \offprints{M. Altmann}

   \institute{Sternwarte der Universit\"at Bonn,
              Auf dem H\"ugel 71, 53121 Bonn, Germany\\
              \email{maltmann,deboer@astro.uni-bonn.de}
         \and
             Dr. Remeis-Sternwarte, Sternwartstr. 7, 96049 Bamberg, Germany\\
             \email{edelmann@sternwarte.uni-erlangen.de}
             }

   {\date{\today}}

   \abstract{ We have analysed the kinematics of a sample 
of 114 hot subdwarf stars. 
For 2/3 of the stars, new proper motions, 
spectroscopic and photometric data are presented. 
The vast majority of the stars
show a kinematic behaviour that is similar to that of Thick Disk stars.
Some  stars
have velocities rather fitting to 
solar, i.e. Thin Disk, kinematics. 
About $\sim$15 objects have orbital velocities
which differ considerably  from those of Disk stars. 
 These are members of the Galactic Halo. 
We investigated the velocity dispersions 
and calculated the orbits. 
Most stars feature
orbits with disk character (eccentricity of less than 0.5), 
a few reach far above the Galactic plane and have very eccentric orbits 
(eccentricity of more than 0.7).
The intermediate eccentricity range is poorly populated. 
This seems to indicate that the (Thick) Disk and the Halo 
are kinematically disjunct. 
Plotting a histogram of the orbit data points along $z$ 
 leads to the $z$-distance probability distribution of the star; 
doing this for the whole sample leads to the $z$-distance probability 
distribution of the sample. 
The logarithmic histogram shows two slopes, 
each representing the scale height of a population. 
The disk component has a scale height of 0.9~($\pm$0.1)~kpc,
which is consistent with earlier results 
and is similar to that of the Thick Disk.
The other slope represents a component with a scale height 
$\sim$7~kpc, 
a much flatter gradient than for the disk component. 
This shows that the vast majority of the sdBs are disk stars, 
but a Halo minority is present, too. 
The kinematic history and population membership of the sdB stars
on the whole is different from that of the cooler HBA stars, 
which are predominantly or even exclusively Halo objects. 
This leads to the question, 
whether the Halo sdB stars are of similar origin as the HBA stars, 
or whether their kinematical behaviour  possibly represents another origin, 
such as infalling stellar aggregates or inner disk events. 
}
   \maketitle
%

\section{Introduction}
\label{intro}
To understand the structure and evolution of galaxies
it is essential to study the gaseous and stellar components of the Milky Way.
No other large galaxy gives us access to the spatial distribution and
kinematics of stars in such detail.
 A vital aspect is the study of older stars, because it brings insight 
into the formation and evolution of the Galaxy and thus of galaxies in general 
(assuming that our Galaxy is typical with no 
or only small peculiarities).

One approach to studying the distribution of stars 
is to conduct a star count over all stars in fields 
at different Galactic latitudes such as done by \citet{Reidmaj1993}.
 Fitting models to account for luminosity class, metallicity, completeness,
distributions and number densities of populations etc. to the raw results 
then leads to scale heights and space densities.
This method relies
on a very large number of stars, 
which means that the statistical errors are small.
On the downside, these studies heavily rely on models, introducing 
uncertainties caused by possibly poorly known input parameters.

Therefore many studies use a certain well defined star type 
as a tracer rather than all available stars. 
In most cases these tracers are evolved stars, which, while
being relatively rare, are bright so that studies using them 
extend deeper into the galaxy. 
Widely used objects are giants and horizontal branch (HB) stars
(especially RR-Lyraes, because they are, as variables very easily identified). 
Nowadays the recent deep surveys enable studies using very low mass 
main sequence stars as tracers as well \citep{CADIS}.

Studies of the spatial distribution of a sample of stars 
give insight into the general structure of the Galaxy, 
e.g. revealing various populations of stars.
Adding kinematical data gives us access to the motions of the stars 
forming these groups. 
Stars belonging to different populations show widely differing 
kinematical behaviour. Some components
of the Milky Way are rapidly rotating with little dispersion in the velocities
of the members while others show only little rotation but high dispersions.
These differences between the populations give us evidence of how these
parts of the Galaxy are formed. Such studies have been conducted for quite 
a variety of different object types, such as high proper motion stars 
\citep{Carney96},
local dwarfs \citep{Schuster97} or globular clusters 
\citep{Din971,Din991,Din992}. The kinematics of sdB stars have been studied
by \citet{Colin94}, \citet{T97} and \citet{B97}.

Blue subdwarf stars, such as sdB stars, are particularly suited for such a 
study because they have spectra that can be analysed with relatively simple
methods. 
 And since their spectra are quite unique, there are no other objects
they can be easily confused with. 

 Horizontal Branch stars (HB) are
core helium burning objects after the first red giant (RGB)
phase. Their appearance depends on the mass of the hydrogen envelope
they retain, while the
He core mass is relatively constant over all types. 

Extreme HB stars (EHB)
such as sdB stars only have a very thin H envelope of less than
0.02 M$_{\odot}$.
HB stars hotter than about 10\,000 K have metal abundances
heavily altered by effects of diffusion and levitation in their stable and 
non-convective atmospheres, as has been found by many studies
\citep{Bonifacio95,Moehler99,Behr99}.
Since the present element abundances do represent
the initial metal content, 
information of the population membership of hot HB stars is 
only accessible through their spatial distribution and kinematics. 

 The study presented here 
is a continuation of the work done by \citet{B97}, increasing the number of 
stars by a factor of almost three, adding some more local stars and stars from
the HE-survey carried out by the Hamburger Sternwarte. 
These are on average a little further away than those
of the PG-survey \citep{PG} used in the earlier work, but on the whole 
at significantly larger $z$-heights as most of them are located 
at higher Galactic latitudes than the PG-stars used in \citet{B97}. 
 With this enlarged sample, one can therefore expect the
probability to include Halo objects in the sample to be larger,
 because their relative density is higher than near the Sun
where disk stars dominate.

Sect.\,\ref{data} deals with the composition of our sample, data acquisition 
and reduction. 
In Sect.\,\ref{samplesel} we discuss possibly selection effects induced by 
the sample composition.
The analysis of the kinematics and orbits is described in Sect.\,\ref{kinorb}, 
the determination of a scale height from the orbits in 
Sect.\,\ref{scaleheight}. 
In Sect.\,\ref{discussion} we discuss our results in a larger context. 
Finally Sect.\,\ref{conclusions} gives our conclusions and an outlook
towards the future.

\section{Data and data reduction}
\label{data}
\subsection{The sample}
\label{sample}
     The sample of 114 sdB/OB stars (sdOB stars are hotter versions of sdB stars with
\teff\ of more than about 32\,000 K featuring some He lines in their spectra. Since
they are - unlike the even hotter sdO stars - 
generally believed to be HB-like stars we use the term sdB stars throughout this work)
is composed of objects taken from 
several sources. 59 stars, located in the southern polar cap (SPC) of our 
galaxy have been taken from the Hamburg/ESO survey (HE).
For these new data have been obtained. 
We further included the 41
 objects published in \citet{B97}, which were mainly taken from the Palomar Green catalogue
\citep{PG}.
17 stars with Hipparcos or Tycho 2 data\footnote{Three objects 
(HD 205805, CD $-$38 222, PG 1519+640) are in common with \citet{B97}. 
We use here the Hipparcos or Tycho2 proper motion data
which was not yet available at the time of publication of \citet{B97}.}
 have also been included as well as one star 
(PG 1716+426) whose kinematics was analysed by \citet{G98}. 

\subsubsection{Selection effects due to sample composition?}
\label{samplesel}

\begin{figure}

   \centering
   \epsfig{file=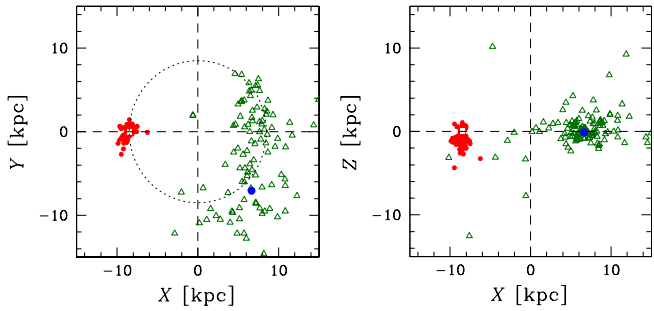,scale=0.45}
   \caption{Distribution of the stars of our sample today (full hexagons)
        and approximately half a revolution (100 Myr) earlier (open triangles)
        showing that our stars, {\it now concentrated in a small volume}, 
        came in fact from all over the Galaxy.
        The left panel shows the distributions on the Galactic plane, the right
panel shows the distributions perpendicular to the plane (along the $X$-axis). 
       The full square and full circle show
       the position of the Sun today and 100 Myr ago respectively.
       The circle on the left panel shows the present galactocentric 
       distance of the Sun.
       The Galactic centre is in the middle of both diagrams, 
     the dashed lines show the zero line for each coordinate. 
               }
   \label{sampledistr.fig}
\end{figure}

The sample of sdB stars discussed in this paper 
is the collection of all relatively nearby stars we could lay our hands on. 
The sole criterion was, that we should have or could obtain, for each sdB star 
its distance, its radial velocity and its proper motion.
We have not aimed at obtaining and working with an 
observationally unbiased sample. 
 To show that this will not cause any problems in the results 
is the goal of the discussion of this section.
For the present positions of the stars of the sample see 
Fig.\,\ref{sampledistr.fig}. 

The main sources of our objects were the catalogues from the surveys  
PG and HE. 
The PG and HE surveys were conducted at high Galactic latitudes. 
These surveys are incomplete at the bright end 
(the PG for stars brighter than $B \sim 12$ mag, 
the HE $\sim 13$ mag)\footnote{These upper limits
are not explicitly stated in \citet{PG} but for the Hamburg survey 
they are stated in the description of the catalogues:  
{\tt http://www.hs.uni-hamburg.de/} 
{\tt EN/For/Eng/Sur/index.html}.}. 
 
Thus nearby stars at high Galactic latitudes may be underrepresented. 
However, since stars presently nearby may venture in time to almost 
any Galactic location (see Fig.\,\ref{sampledistr.fig}), 
our sample must contain sufficient presently distant stars which 
at other times would have been near the Sun  being then that bright, 
that they would have escaped the PG and HE surveys. 

The stars we found in the {\sc Hipparcos} data base,
which are mainly from the SB-survey of blue SGP objects \citep{SB}, 
are relatively bright and thus nearby. 

The catalogues available normally do not (or rarely) contain stars 
presently at {\it low} Galactic latitudes. 
All of those low $b$ stars having Thick Disk or Halo kinematics 
would at other times have been detected in surveys like PG and HE 
(see Fig.\,\ref{sampledistr.fig}). 
Stars with that kind of kinematics 
are thus not underrepresented in our sample. 
Most of the stars presently in the disk (thus missing from our sample) 
and which do have disk kinematics would always have been missed. 
Thus stars with Thin Disk like kinematics are 
probably somewhat underrepresented in our sample. 

Summarising, 
in spite of having used data from various special catalogues 
dealing with particular observational selections of all stars available, 
our sample is only lacking (to an unknown amount) in stars with 
Thin Disk kinematics.

\subsection{Obtaining the data}
\label{dataaq}
 The data has been acquired over the past few years, mainly from the ESO La Silla.
  For the 41 stars
of \citet{B97} we took all of their data, except for a few cases. 
A few proper motions have been taken from the {\sc Hipparcos} catalogue 
(\citealt{HIP}),
some of the other data from various sources in the literature (see Sect. 2.3).

\subsubsection{Spectroscopic data}
\label{dataaqspec}
The spectra of the southern HE-stars
have been taken at La Silla, Chile, with the ESO 1.54m Danish telescope
using the Danish Faint Object Spectrograph and Camera (DFOSC)
covering a spectral range from $3500$-$5500$~\AA\ and with the
ESO 1.52m telescope using the Boller
and Chivens spectrograph covering a spectral range from $3500$-$7000$~\AA.
The spectral resolution is about 5.0 and 5.5~\AA, respectively.
In order to conduct the wavelength calibration, 
especially because radial velocities
were to be derived, an emission lamp spectrum was taken after the object
spectrum, while the telescope was in the same position as during the
object exposure. 
For the flux calibration spectra of the Oke-standard star Feige 110
\citep{Oke1990} were obtained. 
Exposure times were between 120~s and 3600~s, 
depending on the brightness of the object. 

\subsubsection{Astrometric and photometric data}
\label{dataaqphotast}

For each object photometry in $B$ and $V$ is required. For the astrometry
images in bands in the redder part of the spectrum are needed to minimise effects 
from differential refraction (see e.g. \citealt{Brosche89}). Therefore images were
taken in $B$, $V$ and $R$ passbands. In order to calibrate the proper motions to the
extragalactic reference frame, deep and relatively wide field exposures of
the fields surrounding the stars are also needed.

  Data acquisition of CCD images for photometry and as second epoch material 
for the determination of proper motions has been combined wherever possible.

The data was obtained with the 
1.54 m Danish telescope and the DFOSC focal reducer at La Silla in 
January 1999, October 1999 and September 2000. The exposure times 
were between 600 s and 900 s 
($V$ and $R$ only) for those obtained at
La Silla. As the electronic gain of the CCD camera used for the La Silla images
 is such that the target star is overexposed in most 
cases, an additional short exposure was made with an exposure time of 5 to 120~s 
depending on the approximate magnitude of the star. To complete the photometry
a short $B$ exposure was taken as well. On all nights Landolt standard stars
(\citealt{LANDOLT}) were taken at least twice per night, mostly three
times per night.  

\subsection{Data reduction and analysis}
\label{datared}
\subsubsection{Spectroscopy: radial velocities, $\log g$ and $T_{\rm eff}$}
\label{dataredspec}
 The spectra were reduced using 
the ESO-MIDAS package. After bias subtraction and flat field correction
the background was subtracted and the one dimensional spectra were extracted from 
the two dimensional frames.
Thereafter a wavelength calibration was performed
with calibration spectra recorded immediately after each stellar spectrum.
Then all wavelength calibrated spectra were corrected for 
atmospheric extinction using the
extinction coefficients of La Silla \citep{Tug1977}.

The observed spectra (line profiles of Balmer, He I and He II lines)
were then fitted to synthetic spectra
calculated from fully 
line blanketed LTE model atmospheres
\citep{Heber2000} as well as hydrogen and helium blanketed NLTE model 
atmospheres \citep{Napiwotzki1997}. 
The model spectra were convolved to match the
spectrograph's resolution and shifted by the approximate radial velocity.
The fitting was accomplished using the fitting routine of \citet{Lemke1997},
which is based on the procedure of \citet{Bergeron1992} and \citet{Saffer94a}.
The values of the fit parameters \teff, \logg\ and $n$He, the helium abundance, 
will be published in \citet{Edelmann2002}.

Spectroscopic distances were determined using the method of flux conservation.
For this we first obtained the angular stellar diameter by comparing model 
atmospheric fluxes with the dereddened apparent $V$-magnitude. 
From the determined \logg, the radii of the stars are calculated. 
We assume the mass of the sdB stars to be 0.50 \msun. The errors of \logg\ are 
estimated to be 0.1 dex and for \teff\ 2.5\%. The error in the photometry
is about 0.02 mag.
The error of the distances was determined by error propagation 
to be in the order of 10\%.

Finally radial velocities were derived from the centres of gravity 
for all Balmer and helium lines. 

The radial velocity of each star was then calculated by averaging the obtained 
single line velocities.
These then were transformed to heliocentric values.
The resulting radial velocities have an error of about 30 \kms.

Some of the stars show 
signs of companions in their spectra.
These spectroscopically obvious secondary components contribute
significantly to the continuum flux. The routine we employed to
fit spectral models to the spectra critically depends on the local
continuum. Therefore in the case of a binary star the continuum of
the secondary star must be taken care of. As this is a rather
tedious procedure and only a few stars are affected we did not
include these in our final sample. This does not apply for those
stars taken from the literature which were measured using other
fitting techniques, or in the case of SB~744, where the secondary
continuum was subtracted \citep{UB}. 

sdB stars may be the result of close binary evolution.  This means
that even many of those that do not show any sign of binarity 
in their spectrum may have a close companion. 
 The companion star, being too faint to show up in the spectrum 
(through spectral lines or a redder continuum), 
must either be a white dwarf or a low mass main-sequence star.
Such unseen companions do not play a role for
the method of the determination of physical parameters. They
however {\sl do} play an important role for the radial velocity
determination. Especially if the secondary star is a white dwarf
of similar or even higher mass  than that of the sdB primary, 
the measured radial velocity may have a large amplitude -- in some cases over
200 \kms. Indeed, recent studies have found many stars with
variable radial velocities (see e.g. \citealt{Maxted01,Morales2002}
etc.). As far as possible we have used systemic radial
velocities
published in
those studies and some by Marsh (priv. comm.). Unfortunately we
only have access to a few values -- for most of our stars we only
have a single value. Luckily the amplitudes of the majority of
the variable radial velocities are far less dramatic, in the order of
50 \kms\ or less. 
The percentage of close binaries is by no means
certain -- the most recent (Napiwotzki, Edelmann, priv. comm.) 
results show that only about 30\% of the sdB stars do show a variation 
in their radial velocity indeed, in contrast to two thirds 
as estimated in earlier studies. At present we cannot
quantify the influence of variable radial velocities on our
results; we will have to bear this problem in mind, and try to
determine the systemic radial velocities of objects with suspicious
(i.e. very large) values in the future.

\subsubsection{Photometry}
\label{dataredphot}
The basic CCD data reduction, i.e. bias subtraction, flatfield 
correction\footnote{Flat field correction was somewhat cumbersome. There
were noticeably differing large scale structures on the twilight flats
made in the evening and during the morning twilight,
resulting in a residual gradient remaining on the flat field corrected 
images as well. 
We therefore combined the long exposed object frames such that 
all stars could be eliminated. This sum image has the true flat gradient. 
The twilight flat was blurred and the original twilight flat was divided 
by the blurred one to get the small scale structure. 
This was then added to the science-frame flat, 
to be used for the final flat correction.} 
and correction of bad pixels and rows were accomplished using the IRAF
data reduction package which was also used for the subsequent photometric
analysis.

The photometric measurements were conducted using aperture photometry. 
The aperture diameter
used was 14\arcsec, which is the size also used by \citet{LANDOLT} for his
standard star photometry as used to calibrate our measurements.
For those stars, for which we could not derive an accurate magnitude 
(because the night was not photometric) we used values from the literature, and in some cases
if literature values were not available, their $bjdss$-magnitudes.

The majority of our stars is located at intermediate to very high Galactic
latitudes. Nevertheless the photometry must be extinction corrected, to minimise
systematic distance effects.
Since almost all stars of our sample are located at $|z|>200$ pc, thus likely 
above the Galactic dust layer, one can assume that all interstellar extinction
is in front of the star. Therefore we decided to use reddening maps, such as 
those from \citet{Schlegel} or \citet{BH}. The latter have no data for about
30\% of our stars, namely those located near the SGP.

The 
values taken from \citet{Schlegel} are a little larger than those of 
\citet{BH}. Subtracting 0.02 mag
from their values, which the authors suggest to make the data comparable
seems to overdo it. As a good compromise
we decided to take the \citet{Schlegel} reddenings, reduced by 0.01 mag.
The very few resulting negative values were taken as $E_{B-V}=0$.

For the majority of the stars $E_{B-V}$ was
below 0.02 mag, a few had an $E_{B-V}$ of between 0.04 and 0.05. This means
that a residual $E_{B-V}$ will cause an error in the distances which is small
compared to the influences of the other errors. For the stars for which 
we took the data from the literature we also used the extinction 
values from there. 
 
\subsubsection{Astrometry}
\label{dataredast}
\paragraph{1$^{\rm st}$ epoch material:}

The biggest problem encountered when determining proper motions of stars is the
availability of suitable 1$^{\rm st}$ epoch material. 
 Before 1950 photographic plates taken are mostly only of areas 
of special interest with conspicuous objects. 
So one has to rely on whole sky surveys, such as the
Palomar Observatory Sky Survey (POSS), and its southern extension, the
UK-Schmidt Survey. The latter was completed in the 1970's and early 1980's
while the POSS was accomplished between 1948 and 1958. In recent times several
scanned versions of these plates became available, like the DSS or the APM 
(\citealt{APM1}, \citealt{APM2}). 
The APM catalogue has been used for the determination of 
proper motions of objects north of the equator in the past \citep{B97}.
Unfortunately for
$\delta<-15$\degr, the APM relies on recent plates,
rendering it unsuitable as first epoch material. Therefore we took
positions derived from the DSS. 
These scans have an image scale of 1.7\arcsec/pix
and are made from the blue plates of the UK Schmidt-survey.
This could present a problem, 
because blue light is more affected by differential refraction than red light. 
However, the plates were taken
at very low airmass, most of them near the meridian. The declinations of 
our objects are between $-17$\degr\ and $-55$\degr, so they are not more than
25\degr\ off zenith when passing the meridian.

Of the position determining methods generally used to get the plate coordinates,
DAOPHOT is presumably not the best for digitised photographic plates, 
because it uses a PSF, and the PSFs of all photographic stellar images 
are different, in contrast to those of CCD images.
For this reason we decided to use SEXTRACTOR \citep{SEX}. 
Comparing positions
derived using DAOPHOT and SEXTRACTOR shows a good agreement for well exposed
stars, with $\sigma(\Delta x,\Delta y)\simeq$ 0.02 pix, or 0.035\arcsec.
In order to compare
DSS and APM positions, we derived proper motions for 9
stars taken from a field north of the equator using both sources of first epoch material. The
standard deviations of the differences between the pairs of proper motions
was $\sim 2.5$ mas/yr. So we conclude that the error of the proper motions
of the southern stars, which have a much smaller epoch difference,
is about 4 - 5 mas/yr. Adding the error of the galaxy calibration, which is in
the order of 1 - 2 mas/yr, the proper motions derived from DSS first epoch
material are good to about 5 - 6 mas/yr. 

\begin{table*}
\setlength{\tabcolsep}{1.4mm}
\caption[]{The observational data of the stars comprising our sample. The $B-V$, $V-R$ and $E
_{B-V}$ values are only listed for
those stars with new data.}
\begin{tabular}{lllrrrrrrrrrr}
\hline
Name & \multicolumn{2}{c}{$\alpha (2000.0) \delta$} & $V$ & $B-V$ & $V-R$ & $E_{B-V}$ & $d$ &
 \multicolumn{1}{r}{$\mu_{\alpha}\cos\delta$} &
\multicolumn{1}{c}{$\mu_\delta$} & $v_{\rm rad}$ & \multicolumn{1}{c}{Source$^a$} \\
  & [$^{\rm h\; m\; s}$] & [$\degr\; \arcmin\; \arcsec$] & \multicolumn{4}{c}{[mag]} & [kpc]
& \multicolumn{2}{c}{[mas/yr]} &
  [\kms] &  \\

\hline
\input{h4542t1a.tab}
\hline
\end{tabular}
\label{alldata.tab}
\end{table*}
\addtocounter{table}{-1}
\begin{table*}
\setlength{\tabcolsep}{1.4mm}
\caption[]{The observational data (cont.)}
\begin{tabular}{lllrrrrrrrrr}
\hline
Name & \multicolumn{2}{c}{$\alpha (2000.0) \delta$} & $V$ & $B-V$ & $V-R$ & $E_{B-V}$ & $d$ &
 \multicolumn{1}{r}{$\mu_{\alpha}\cos\delta$} &
\multicolumn{1}{c}{$\mu_\delta$} & $v_{\rm rad}$ & \multicolumn{1}{c}{Source$^a$} \\
  & [$^{\rm h\; m\; s}$] & [$\degr\; \arcmin\; \arcsec$] & \multicolumn{4}{c}{[mag]} & [kpc]
& \multicolumn{2}{c}{[mas/yr]} &
  [\kms] & \\

\hline
\input{h4542t1b.tab}
\hline
\end{tabular}

{\small $^a$: Data sources: unmarked, data from this work and from \citet{Edelmann2002}, B+97=\citet{B97},
AB00=\citet{AB2000}, HIP=\citet{HIP}.
Additional sources: Photometry: HE=Hamburg-Eso-Survey, K+99=\citet{Koen99}, 
B+92=\citet{Beers92}, L+00=\citet{MCTS2000}.
 Spectroscopy (\logg, \teff, $v_{\rm rad}$): TM= Marsh (priv comm), 
M+99=\citet{Moran1999}, M+02=\citet{Morales2002},
PI=\citet{P1}, KHD=\citet{Kilkenny2}, UB=\citet{UB}.
4. Astrometry: TYC=\citet{Tycho2}, G98=\citet{G98}\\
$^b$: \citet{B97}, but revised values for the proper motions.}
\end{table*}

\paragraph{2$^{\rm nd}$ epoch material:}

As second epoch material, the same CCD-exposures were used as for the 
photometry (see Sect. 2.2.2). The DFOSC data has an image scale 
of 0.39\arcsec/pix.
For the determination of plate coordinates,
SEXTRACTOR was used for the CCD-images as well. 

\paragraph{Reference catalogues:}

 The third item required for the derivation of proper motions is a reference 
catalogue of the field preferably including proper motions of its stars. 
However,
for small-field astrometry, e.g. with CCD data, there are not enough stars with
proper motions in a typical field. 
 Even the Tycho-2 catalogue provides only about
1~to~5~entries in a DFOSC field of 13.7\arcmin$\times$13.7\arcmin,
almost all of which are saturated even on the shortest exposures. 
 This means that we had to rely on catalogues without proper motions 
and had to accept that the resulting proper motions are relative ones,
which have to be calibrated to an absolute reference frame. We accomplish this
by using the background galaxies in the field which have 
by definition a zero proper motion (see below).   

We created a reference catalogue based on our CCD-data using the 
USNO-A2.0 catalogue \citep{USNO} as start catalogue. 

\paragraph{Reduction:}

The astrometric reduction was performed using the BAP-software
of Michael Geffert (see e.g. \citealt{G97}). In a two step iteration we first
determined celestial positions for each plate/CCD-frame, which were then 
averaged to give a catalogue with positions and proper motions.
This catalogue was then in turn used as new reference catalogue for the
second iteration. For the plate reduction we used a plate model with 3${\rm rd}$
order terms.

The resulting errors (based alone on the second epoch, since there is only one
first epoch position)
 are in the
order of 1 mas/yr with only a few stars having an error larger than 2 mas/yr. 

\paragraph{Calibration:}

As indicated above, the resulting proper motions are relative only.
 Therefore we transfer
them to an absolute reference frame using the background galaxies. For this
one determines the apparent galaxy proper motion centroid and subtracts this from
the stellar proper motions. The galaxies are identified and separated from
stars by criteria of FWHM of their images, SEXTRACTORS stellarity index and
additional user defined selections to account for the larger error of FWHM for
faint objects. This method is well established and often used in galaxy
searches. For each star we found between 15 and 120 galaxies,
mostly between 40 and 70. The limiting factor is the 1$^{\rm st}$ epoch material;
 the long CCD exposures, which were used for galaxy searching, usually yielded
several hundred galaxies.

Unfortunately most background galaxies are 
very faint, show too much intrinsic (spectral dependent) structure, 
and are generally undersampled in the digitised photographic plates. 
Therefore the derived proper motions scatter considerably. 
To get rid of outliers we expunged objects deviating by more than 2 $\sigma$. 
This gave us in general a standard deviation of 10 to 15 mas/yr and 
a resulting error of 0.5 to 2.5 mas/yr in almost all cases. 
Such standard deviations are not unusual, 
given the difficulties described;
\citet{O94b} arrived at values of 8 mas/yr for their standard deviation
of the galaxy proper motions, which is consistent with our value, 
since their epoch difference is more than twice ours.
The errors are almost as large as the values in all but a few cases. 
We nevertheless applied the galaxy calibration to put the proper motion in
the extragalactic reference frame. 

\subsection{The final sample}

The final sample consists of 114 sdB/OB stars. The observational data and data derived from observations, 
such as positions, $V$-magnitudes, distances, proper motions and radial velocities
 are shown in Table\,\ref{alldata.tab}. This applies to all data, the new data and older data from the literature.
The values of \logg\ and \teff will be published in \citet{Edelmann2002}.

\section{Kinematics and orbits}
\label{kinorb}
\subsection{Calculating velocities and orbits}
\label{kinorbcal}
The observational quantities $\alpha,\delta,d,\mu_{\alpha},\mu_{\delta},
v_{\rm rad}$ - shown in Table\,\ref{alldata.tab} - are transformed into the $X,Y,Z,U,V,W$ system (for details, see
\citealt{AB2000}, \citealt{B97}). Additionally the orbital velocities
$\Theta$ (the velocity component in the direction of Galactic rotation 
projected to the Galactic plane)
and the velocities towards the Galactic centre $\Phi$, kinetic
energies and angular momenta are calculated\footnote{Because all of the stars 
are local ($d < 5$ kpc), $U$ and $W$ are quite similar to $\Phi$ and $\Theta$;
however, especially for stars being well away from $Y=0$~kpc, $\Theta$ and 
$\Phi$ become linear combinations of $U$ and $V$.}. The resulting values for each star are 
represented in Table\,\ref{allgaldat.tab}. 

Furthermore we calculated orbits for the stars of our sample using the
Galactic gravitational potential model of \citet{AL91} backwards in time over
10 Gyr in steps of 1 Myr (for more details see \citealt{AB2000}).

From the shape of the orbits we derived the apo- and perigalactic
distances, $R_{\rm a}$ and $R_{\rm p}$,
 and the eccentricity ($ecc$) given by
\begin{equation}
ecc=\frac{R_{\rm a}-R_{\rm p}}{R_{\rm a}+R_{\rm p}}
\end{equation}
We also wish to consider the maximum distance a star reaches from
the Galactic plane, $z_{\rm max}$. However, since the
gravitational potential diminishes at larger galactocentric
distance, $\varpi$, we calculated

\begin{equation}
nze=\frac{z_{\rm max}}{\varpi(z_{\rm max})},
\end{equation}
the normalised $z$-extent of the orbit,
which is more relevant than
$z_{\rm max}$ alone. The eccentricities and $nze$ are also shown in Table\,\ref{allgaldat.tab}.

Finally we stress that we calculated
the orbits for the long timespan of 10 Gyr only because it gives a better
representation of the orbits' shapes. This applies in particular
to low velocity Halo type stars taking a long time to complete one revolution.
The reader should be aware that orbits are subject to gradual change 
over time due to gravitational interactions with local high density areas 
in the Galaxy. 
Therefore the orbit calculated over a certain time does not represent
the true trajectory in the far past.

 The errors of the velocity components depend on the uncertainties
of the input values, namely 10~\% for the distances, 30~\kms\ for the radial velocities and 5~mas/yr
for the proper motions (which means an error in tangential velocity of about 30~\kms\ at 1.5 kpc).
For a star at a typical distance of 1.5~kpc this results in an error estimate of 35~\kms.
Stars at lower distance have a smaller error in the tangential velocity while objects further away have larger errors. 

Those stars with Hipparcos data, or with available systemic radial velocities have far smaller errors than indicated in this
error estimation. In contrast to this, stars in undetected close binary systems (see Sect.\,\ref{dataredspec}), could have
far more uncertain radial velocities. 

\begin{table*}
\setlength{\tabcolsep}{1.35mm}
\caption[Positions, velocities and morphological data of all stars]
{Positions, velocities (given in the Galactic euclidic system $XYZ\,UVW$ and
$\Phi,\Theta $) as well as the angular momentum $I_z$ and
morphological orbital data ($R_{\rm a}$, $R_{\rm p}$, $z_{\rm max}$ and eccentricity
($ecc$), normalised $z$-extent ($nze$), see text) of all stars. }
\begin{tabular}{rlrrrrrrrrrrrrrr}
\hline
Name & {\it X}& {\it Y} & {\it Z} & {\it U}  & {\it V} & {\it W} &{$\Phi$}& {$\Theta$} & {$I_z$} & $R_{\rm a}$ & $R_{\rm p}$ & $z_{\rm max}$ & $nze$ & $ecc$ \\
& \multicolumn{3}{c}{[kpc]} & \multicolumn{3}{c}{[km\,s$^{-1}$]} &\multicolumn{2}{c}{[km\,s$^{-1}$]} & \multicolumn{1}{c}{[kpc\,km\,s$^{-1}$]} & \multicolumn{3}{c}{[kpc]} & & \\
\hline
\input{h4542t2a.tab}
\hline
\label{allgaldat.tab}
\end{tabular}
\end{table*}
\addtocounter{table}{-1}
\begin{table*}
\setlength{\tabcolsep}{1.35mm}
\caption[Positions, velocities and morphological data of all stars]
{Positions, velocities and morphological data of all stars (cont.)}
\begin{tabular}{rlrrrrrrrrrrrrrr}
\hline
Name & {\it X}& {\it Y} & {\it Z} & {\it U}  & {\it V} & {\it W} &{$\Phi$}& {$\Theta$} & {$I_z$} & $R_{\rm a}$ & $R_{\rm p}$ & $z_{\rm max}$ & $nze$ & $ecc$ \\
& \multicolumn{3}{c}{[kpc]} & \multicolumn{3}{c}{[km\,s$^{-1}$]}  &\multicolumn{2}{c}{[km\,s$^{-1}$]} & \multicolumn{1}{c}{[kpc\,km\,s$^{-1}$]} & \multicolumn{3}{c}{[kpc]} & & \\
\hline
\input{h4542t2b.tab}
\hline
\end{tabular}
\end{table*}

\begin{table*}
\setlength{\tabcolsep}{1.5mm}
\caption{$UVW,\Theta\Phi$ velocities, angular momentum, eccentricities and $nze$
with their dispersions for the 110 star sample.}
\label{vel.tab}
\begin{tabular}{p{2.5cm}rrrrrrrrrrrrrrrrr}
\hline
\\[-8pt]
 Subsample & $N$ &  $\bar{U}$ & $\sigma_U$ & $\bar{V}$ & $\sigma_V$ & $\bar{W}$ & $\sigma_W$
& $\bar{\Theta}$ & $\sigma_\Theta$ & $\bar{\Phi}$ & $\sigma_{\Phi}$ & $\bar{I_{\rm z}}$ &
$\sigma_{I_{\rm z}}$ & $\bar{ecc}$ & $\sigma_{ecc}$ & $\bar{nze}$ & $\sigma_{nze}$\\
  &    & \multicolumn{10}{c}{[km\,s$^{-1}$]}& \multicolumn{2}{c}{[kpc\,km\,s$^{-1}$]} & & & & \\
\hline
all         & 114 &  $-$8 &  74 &+198 &
  79 &+12 & 64 &+198 &  80 & +6 &  74 &
+1700 &  705 & 0.33 & 0.22 & 0.51 & 2.74 \\
$R<8.5$ kpc &  52 & $-$13 &  60 & +208 &
 61 & +10& 53 & +208 &  61 & +22 &  62 &
+1698 &  465 & 0.28 & 0.18 & 0.18 & 0.42 \\
$R>8.5$ kpc &  62 &  $-$3 &  70 & +196 &
  84 & +13 & 72 & +196 &  85 & $-$7   &  79 &
+1761 &  902 & 0.35 & 0.23 & 0.30 & 0.49 \\
$ecc<0.55$   &  99 &  $-$5 &  59 & +198 &
52 & +10 & 55 & +199 &  53 &  +5 &  58 &
+1707 &  460 & 0.26 & 0.13 & 0.19 & 0.36 \\
$ecc>0.55$   &  15 & $-$29 & 136 & +196 &
172 & +31 &102 & +193 & 68  & +15   & 138 &
+1655 & 1947 & 0.78 & 0.13 & 2.57 & 7.15 \\
$Z<0.25$ kpc & 10 &  +23 & 46 & +225 & 12 &
 0 & 24 & +225 & 12 & $-$23  & 47 &
+1909 &  97 & 0.15 & 0.09 & 0.03 & 0.03 \\
$Z>0.25$ kpc & 104 & $-$11 & 75 & +195 & 82 &
+14 & 66 & +195 & 84 & +9  & 75 &
+1680 & 738 & 0.34 & 0.22 & 0.55 & 2.86 \\
\hline
Sun          &  1 &  +10 & $-$ & +235 &
 $-$ & +8 & $-$ & +235 & $-$ & $-$10 & $-$ &
+1998 & $-$ & 0.08 & $-$ & 0.01 & $-$ \\
  \hline
\noalign{\smallskip}
\hline
\noalign{\smallskip}
\end{tabular}
\end{table*}

\subsection{Analysis of the velocities and velocity dispersions}
\label{anavel}

\begin{figure}

   \centering
   \epsfig{file=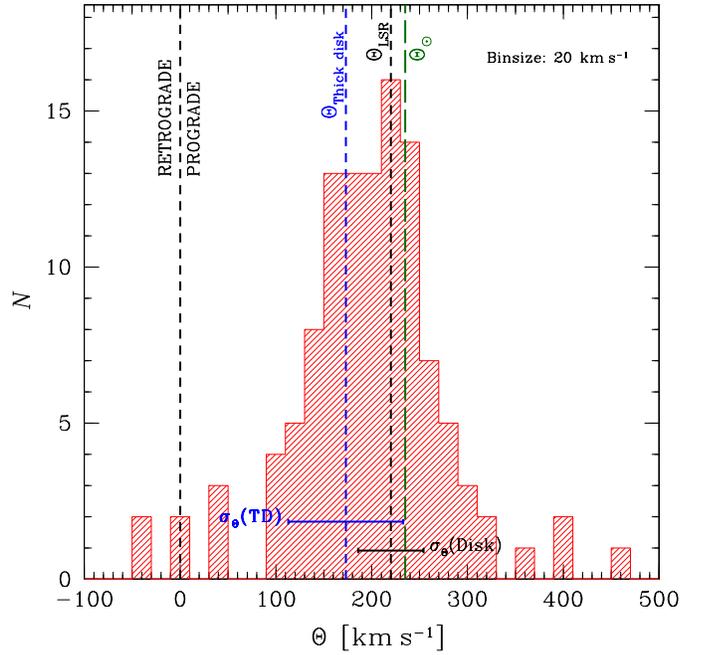,scale=0.45}
   \caption{Histogram of the orbital velocities $\Theta$ for all 110 stars
            of the sample. The values for $\Theta_{\rm Thick\; disk}$
            and $\sigma_{\Theta}{\rm (TD,Disk)}$ have been taken from
            \citet{O94b}.
               }
   \label{velhist.fig}
\end{figure}

For the majority of the objects, the orbital velocities have values which
are similar to those of disk stars.
 This has  been found before by other studies, e.g.
\citet{B97} or \citet{T97}. However a minor portion of the sample has
orbital velocities that are significantly below or above those expected for
disk stars (see Fig.\,\ref{velhist.fig}); some stars have even a near zero or even slightly negative
orbital velocity.
As seen in earlier studies, the majority of the
sdB stars have velocities rather indicative of disk orbits, but with
$\bar{\Theta}$ somewhat lower than that for Thin Disk stars, while the velocity
dispersions are larger (see Table\,\ref{vel.tab}).

 $\sigma_V$ is larger than
expected for the Thick Disk alone, reflecting the fact that the
Halo and Thin Disk components are present in our sdB sample.
A significant old Thin Disk contribution can also be seen by looking at
$\bar{\Theta}$ which is higher than what most studies of Thick Disk kinematics
arrive at (see e.g. \citealt{O94b}). However, due to the composition of the
sample, which is certainly lacking stars currently located at
low $z$-heights, we miss a fraction of the Thin Disk stars (see \ref{samplesel}).

\begin{figure}
   \centering
   \epsfig{file=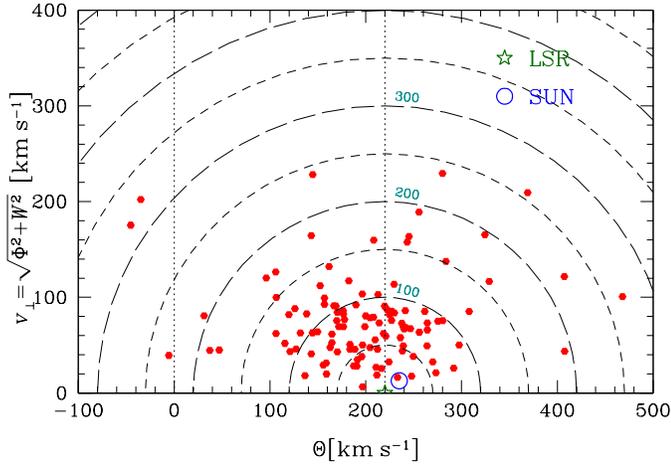,scale=0.45}
   \caption{ Toomre diagram ($\Theta$ versus velocity perpendicular to the
Galactic plane) showing the kinematic divergence of the stars of our sample.
The circles with values in km\,s$^{-1}$ indicate
$v_{\rm pec}=\sqrt{\Phi^2+W^2+(\Theta-\Theta_{\rm LSR})^2}$.
Note the asymmetry in the central condensation of data points
with $v_{\rm pec} < 150$ km\,s$^{-1}$.
A star denotes the LSR and a circle the Sun's values.
           }
   \label{toomre.fig}
\end{figure}

The stars with large $\Theta$ (between 250 and 300~km\,s$^{-1}$) are 
also evident in the Toomre diagram (Fig.\,\ref{toomre.fig}), 
which shows the kinematic divergence of a sample of stars. 
In our case most data points are located within 
$v_{\rm pec}\le~100$~km\,s$^{-1}$, with
the region of $v_{\rm pec}\le 150$~km\,s$^{-1}$ also well populated for
$\Theta \le \Theta_{\rm LSR}$. 
This obvious asymmetry  
shows a behaviour usually known as asymmetric drift. 
It means that kinematically hotter populations tend to rotate slower
than kinematically cooler populations. 
The reason for this effect lies in the greater eccentricity of the orbits 
of such objects and will be discussed in greater detail in the analysis of 
the kinematic behaviour over the whole orbit (Sect.\,\ref{orbvelana}).
A few points lie further out, indicating a kinematic
behaviour quite different from the rest of our sdB stars. The central
concentration is well filled to $\Theta\simeq$ 300~km\,s$^{-1}$.
At very low peculiar velocities ($v_{\rm pec}\simeq$ 30~km\,s$^{-1}$)
only relatively few points are present.

\begin{figure}
   \centering
   \epsfig{file=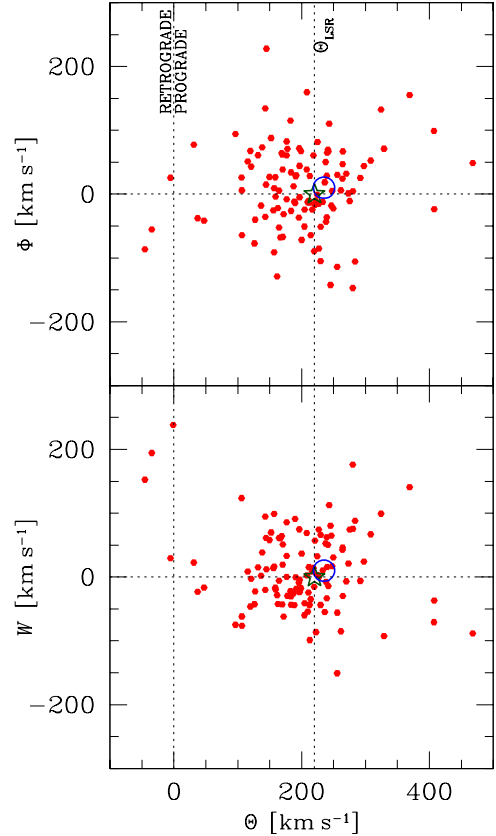,scale=0.60}
   \caption{ Bottlinger and $\Theta-W$ diagrams of the stars of our
sample show the orbital velocity plotted against the two other components,
the velocity towards the Galactic centre ($\Phi$, top) and
perpendicular to the Galactic plane ($W$, bottom).
A star denotes the LSR and a circle the Sun's values.
               }
   \label{bottlinger.fig}
\end{figure}

 The Bottlinger diagrams (Fig.\,\ref{bottlinger.fig})
show a concentration of stars at low values of $\Phi$ and $W$ 
respectively near $\Theta_{\rm LSR}$. However the concentration of data points 
does appear to be slightly shifted in respect to the $\Phi$,$W$=0 axis.
In both panels (but especially in the one showing $W,\Theta$) the points seem 
to be somewhat inhomogeneously distributed.
Again a few stars deviate from the general concentration by a large degree.

The stars of our sample show a behaviour which is kinematically
hotter than but not too different from that of the Sun. This
implies that the majority of our stars belong to the Thick Disk
because their orbital velocities are somewhat lower than those of
stars with solar kinematics. A few stars have orbital velocities
which differ a lot from those of the rest, either being far higher
or lower than those of the rest. Two or perhaps three stars even have
mildly retrograde orbits. These are presumably not disk stars but
members of a non-rotating Halo population. Whether the stars
having a high velocity and those with a low $\Theta$ are of
similar or different origin will be discussed in the next section,
when the orbits are examined. This also applies to those disk
stars which also have relatively high $\Theta$ values.

In order to find differences in kinematics in parts of our sample and to make a crude
populationary separation we applied simple cuts to the sample. 
The normal way to achieve this is by using a non kinematic selection criterion such
as metallicity which is not possible in our case (see Sect. 1). 
For this reason we have to
use kinematical criteria which unfortunately means introducing biases.
First of all we divided our sample with simple criteria 
(Table\,\ref{vel.tab}).
Cuts were made at $R=8.5$ kpc, $ecc=0.55$ and $z=0.25$ kpc, to roughly
 separate Thin Disk, Thick Disk
or Halo. However all populations samples created with these simple cuts 
still show contamination of other populations. The two Disk components could 
not be reliably 
separated, while Disk and Halo seem to be generally well separated.

\subsection{The orbits}
\label{orbits}

\subsubsection{Orbit morphology}
\label{orbmorphana}
\begin{figure*}

   \centering
   \epsfig{file=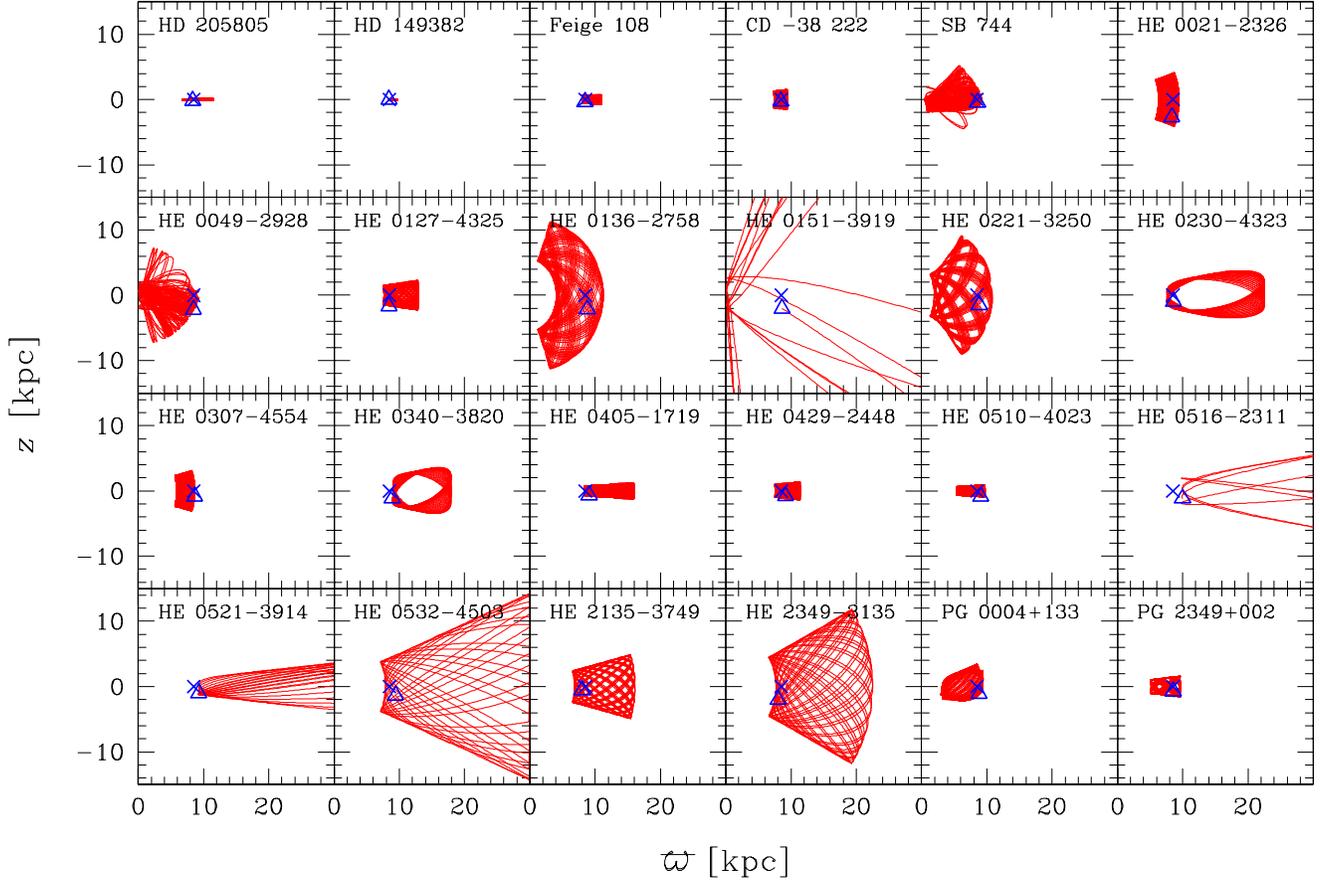,scale=0.92}
   \caption{Examples of orbits of sdB stars. The orbits
    are depicted as meridional plots and calculated over a timespan of 10 Gyr.
    The triangle denotes the current position of each star
   and the cross the current position of the Sun.
   Note that this figure does not attempt to be representative of the sample but to show
   all types of orbits of our stars. Stars with Halo type orbits are
   overrepresented in this plot.
   For the orbit of the Sun (being a Thin Disk star) see Fig.\,1 of \citet{B97}.
        }
   \label{orbits.fig}
\end{figure*}

The morphologies of the orbits show large varieties and we show 
some exemplary orbits in Fig.\,\ref{orbits.fig}. However 
most show box type orbits typical of Disk/Thick Disk stars. 
Six stars have chaotic
orbits, or semi chaotic orbits. These are the stars venturing very close to
the Galactic centre, like the majority of the HBA stars of \citet{AB2000}.
About the same number of stars have orbits going to very large galactocentric
distances, one having its apogalacticon at $\sim$160~kpc. 
These are the stars which have an orbital velocity much higher than that
of the LSR, in some cases approaching the escape velocity of the Milky Way.   

The stars have orbits with eccentricities (see Fig.\,\ref{eccnzehist.fig}, 
left panel) spanning almost the complete
range, however more than 80\% have $ecc<0.5$. This is the region
mainly populated by Disk and Thick Disk stars. A minority has
$ecc>0.7$ with the intermediate zone somewhat underpopulated. This
might mean that the sdBs are part of two kinematically quite
distinct groups, namely one with orbits of small to medium
eccentricity and another having very eccentric orbits.

The distribution of the normalised $z$-extents of our sample
(see Fig.\,\ref{eccnzehist.fig}, right panel) is more peaked than
that of the values for $ecc$,
almost all of the stars having $nze\le~0.4$, 
but there is a long tail to high values.
We do not see a separation at intermediate $nze$ values.
However this is expected as $nze$ is, amongst others, 
a measure for the inclination of an orbit. 
A group of stars all having small orbit inclinations
will show up as a large peak at low values of $nze$ in a histogram. 
In contrast to that, a group of stars having orbits with more random 
inclinations will populate the range of $nze$ without any preferential value 
(except in the case of stellar streams or moving groups). 
This is basically what can be seen in Fig.\,\ref{eccnzehist.fig},
namely a peak of a population of low $nze$ stars and a level distribution of a
group of stars with a large spread in orbit inclination.

\subsubsection{Analysis of the kinematics over the whole orbits}
\label{orbvelana}

The analysis of the current velocities
(Sect.\,\ref{anavel}) gives important information about
the kinematical behaviour and population membership of the sample.
However {\it velocities} are in general {\it not} conserved quantities,
and therefore change over time. Analysing a sample of stars solves
this problem because mean values of the velocity components and
their dispersions are conserved quantities. However they possibly
suffer from selection effects. For example stars have, during a
revolution around the Galactic centre, a considerable range in
orbital velocity $\Theta$ (e.g. the Sun has a $\Theta_{\rm min}$
of 203~km\,s$^{-1}$ and a $\Theta_{\rm max}$ of 237~km\,s$^{-1}$).

 Furthermore,
all stars spend most of the time at galactocentric distances ($\varpi$) 
near the turning points of the orbit, i.e. near the 
peri- and especially apogalactic distance.
 The second Keplerian law and the stellar density law of our Galaxy then
leads to the result that the high velocity part of a sample is less populated
than the low velocity end. 
For a sample of stars this means that the distribution of time related 
velocities is not necessarily a Gaussian but may be a broader,
possibly even a bimodal distribution. 

\begin{figure}

   \centering
   \epsfig{file=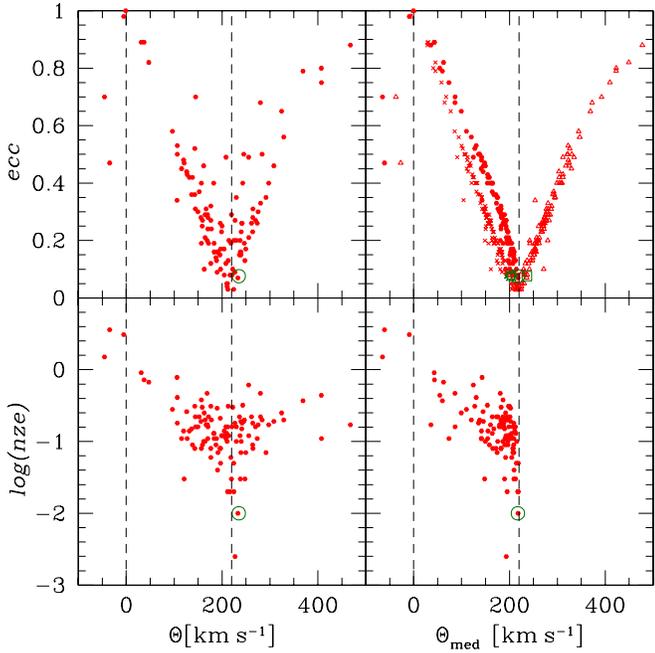,scale=0.45}
   \caption{ Plots $\Theta$ against $ecc$ and $\log nze$:
     On the {\bf left} side the current values of $\Theta$ are plotted,
   on the {\bf right} side the median $\Theta$.
     In the upper right panel the maximum (open triangles)
   and minimum (crosses) $\Theta$ in the orbit are plotted as well.
       The Sun's values are represented by the open circles and
   in the upper right panel as the open square and star.
   $\log nze$ is used rather than $nze$ to gain access to more
   detail at low values.
   The dashed lines indicate the border between prograde and
   retrograde motion and the $\Theta_{\rm LSR}$ at 220 km\,s$^{-1}$.
          }
   \label{nzeecctheta.fig}
\end{figure}

Another consequence of the second Keplerian law is the asymmetric drift, 
i.e. the effect that 
a sample of stars generally stays behind the local standard of rest. 
This lagging behind depends on the eccentricity of a star's orbit;
the more eccentric it is the more the mean orbital velocity deviates 
from $\Theta_{\rm LSR}$, an effect
clearly seen in Fig.\,\ref{nzeecctheta.fig}.
For this reason a sample of  stars with moderately eccentric orbits
is on average slower in Galactic rotation than a sample of stars
with near circular orbits. 
The Toomre diagram (Fig.\,\ref{toomre.fig}) shows this effect especially
well in the asymmetric distribution of data points, with more objects being at
 $\Theta<\Theta_{\rm LSR}$ than at $\Theta>\Theta_{\rm LSR}$.

In Sect.\,\ref{anavel} we noted that there are several
stars with relatively high
orbital velocities ($\Theta~\ge~250$~km\,s$^{-1}$).
 Looking at the values of $\Theta$ along the complete orbits 
of these stars shows that the kinematics is very similar to that 
of stars having $\Theta$ of less than 200~km\,s$^{-1}$. 
This means that every
sample of Thick Disk stars must have a number of objects having a
$\Theta$ significantly faster than the LSR (for solar like orbits
the range in $\Theta$ covered is much smaller). Now looking again
at Fig.\,\ref{velhist.fig}, one sees that the histogram peaks at
$\sim$220~km\,s$^{-1}$ and has a plateau down to 150~km\,s$^{-1}$.
The reason for this could be that the stars come from two
populations, namely a kinematically hotter one, i.e. the Thick
Disk, and a minority of stars, having much tighter orbits, 
representing the Thin Disk.

\begin{figure}

   \centering
   \epsfig{file=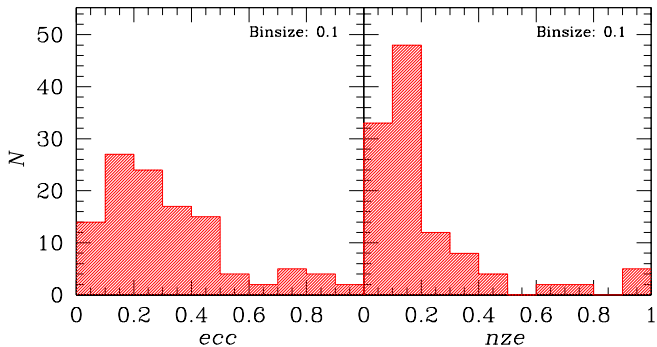,scale=0.45}
  \caption{Histograms showing the distributions of $ecc$ (left panel) and $nze$
           (right panel) for the stars of the sample. Note the peaks at low
           values in $ecc$ and $nze$ and the local minimum in the distribution
           of the eccentricities near $ecc$=0.55.
          }
   \label{eccnzehist.fig}
\end{figure}

In Fig.\,\ref{nzeecctheta.fig} we plotted the current and median
values of the orbital velocities $\Theta$ against the
eccentricities and normalised $z$-extents of the stars. In the
upper left panel of Fig.\,\ref{nzeecctheta.fig} one can see that
most of the stars have low to moderate eccentricity orbits but a
group has high $ecc$ values with a less populated region near
$ecc=0.6$ (which was already evident in the histogram of 
Fig.\,\ref{eccnzehist.fig}). Moreover it is apparent that while there
are more stars having a $\Theta$ smaller than $\Theta_{\rm LSR}$,
about 1/4 of the objects have higher velocities. 
 As $\Theta$ of the stars changes over time we plotted
the median of $\Theta$ over the whole orbit\footnote{We took
 the median rather than the mean because it separates 
the values a little more, and for all except the highly eccentric orbits 
there is only a minor difference between both values.}. 

To show the complete
range of variation in $\Theta$ we also plotted the maximum and
minimum values in the upper right panel of Fig.\,\ref{nzeecctheta.fig}. 
The $\Theta_{\rm med}$ all lie on a line,
except those of the retrograde orbits. Some deviate a little from
this line. These are stars on somewhat more inclined orbits having
a higher $W$ velocity component (which also adds to the velocity
supporting the orbit; a star on a higher inclined orbit can have a
less eccentric orbit with the same orbital velocity, because of
its larger $W$ velocity component). Again, the two groups of stars
can be seen as well as the division at $ecc=0.55$. At $ecc < 0.2$
the trend in $\Theta_{\rm med}$ is small, and at higher values it
gets more pronounced. Again the Thin Disk part, if it exists,
makes up a large part of this low $ecc$ group.

In the lower two panels of Fig.\,\ref{nzeecctheta.fig}
plots of $\log nze$ ($\log nze$ because
most values of $nze$ are clumped together at low values)  against
$\Theta,\Theta_{\rm med}$ are shown.
The left panel shows the distribution of $nze$ of our sample.
The central condensation marks the majority of disk stars,
and the outliers towards low or very
high $\Theta$ are the Halo stars.
The spur towards low $nze$ at LSR velocities consists of
datapoints of solar type orbits.

\begin{figure}

   \centering
   \epsfig{file=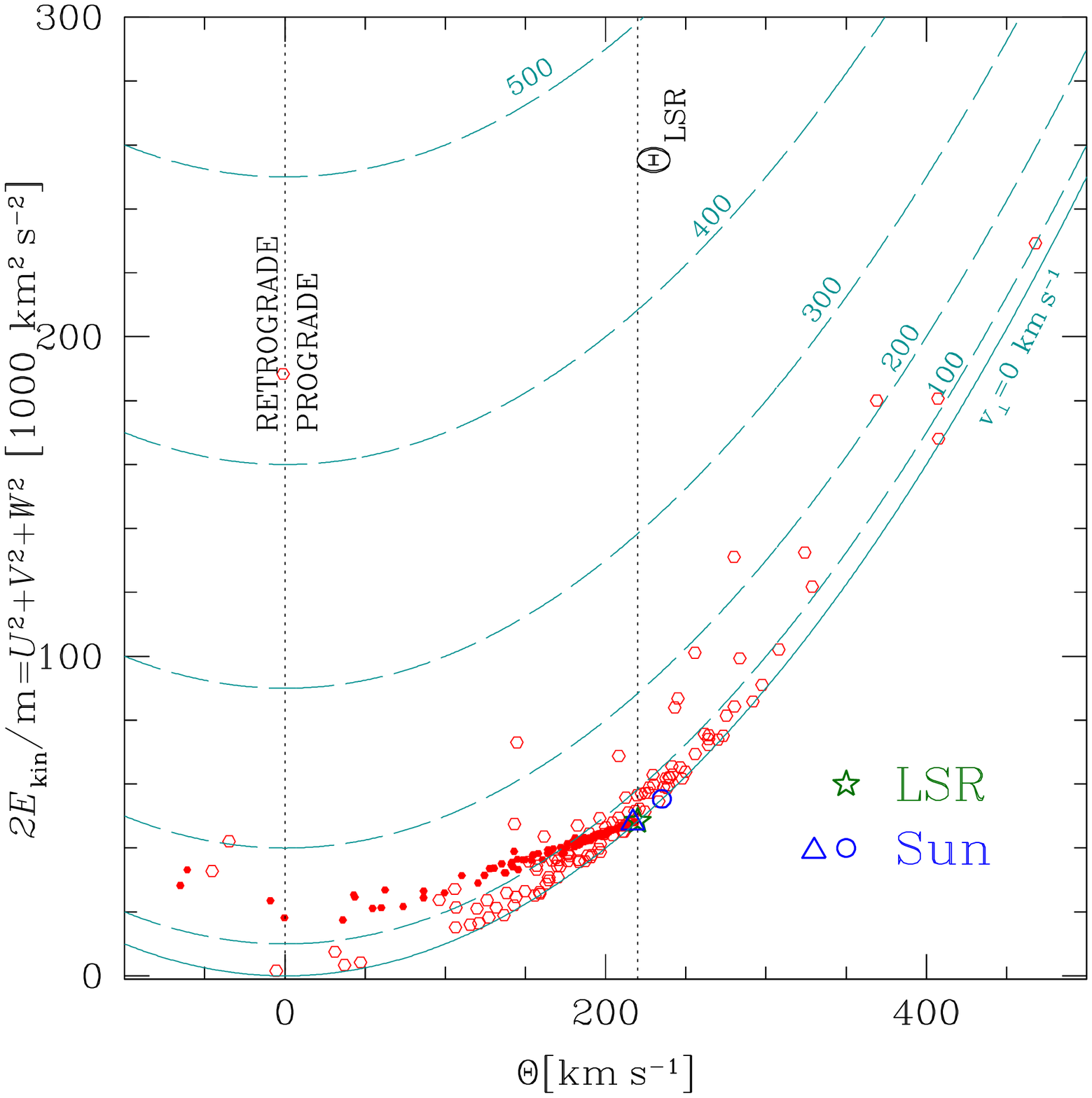,scale=0.45}
   \caption{ Diagram of $\Theta$ as well as $\Theta_{\rm med}$
 against the total kinetic energy ($E_{\rm kin}$).
   Open symbols represent current values of $\Theta$ and $E_{\rm kin}$,
  smaller filled hexagons medianised values
  ($\Theta_{\rm med}$, $E_{\rm kin,med}$).
  The parabolic curves denote lines of equal orthogonal velocity ($v_\perp$,
  the velocity perpendicular to $\Theta$).
  Most of the stars cluster around the LSR along the low $v_\perp$.
  The medianised values lie on an almost straight line pointing
  from the LSR towards the lower left.
  The LSR is marked by a star, the Sun's current values by a circle,
  its medianised one by a triangle.
          }
   \label{Etheta.fig}
\end{figure}

 On the plot $\log nze$ versus $\Theta_{\rm med}$ the different
subdivisions of our sample can be seen even better,
on the left are the Halo stars, which have
a low $\Theta_{\rm med}$ 
(the high velocity Halo stars lie in the far left of this diagram).
The bulk of the Thick
Disk stars cluster around $\Theta_{\rm med}=200$~km\,s$^{-1}$,
$\log nze=-0.9$, and the solar
 kinematics extension protrudes towards the bottom of the diagram.
At values of $\log~nze>-0.7$ there seems to be a gap appearing
between $\Theta_{\rm LSR}$ and the data points. It is possible
that the Thin Disk population represented by the spur more or less
ends at this point (which corresponds to a $z_{\rm max}$ of
$\sim1.7$~kpc or 5~-~6 Thin Disk scale heights). The stars having a
$\Theta_{\rm med}$ of more than about 190~km\,s$^{-1}$ as well as 
$nze$ of less than $\sim$0.2~($\log nze=-0.7$) are the prime
candidates for the Thin Disk component. 

 So potentially
the lower right plot of Fig.\,\ref{nzeecctheta.fig} is a suitable
tool to kinematically separate Thin and Thick disk stars, if these
populations are kinematically detached at all.

Another method of analysing the kinematics of stars is using the
kinetic energy (or total velocity). In Fig.\,\ref{Etheta.fig} we
plotted the kinetic energy \mbox{$2\cdot E_{\rm
kin}/m=U^2+V^2+W^2$} against $\Theta$. The parabolas plotted in
Fig.\,\ref{Etheta.fig} are lines of equal $v_\perp$ (velocity
orthogonal to $\Theta$). 
 This velocity is, together with $\Theta-\Theta_{\rm LSR}$ a measure 
of kinetic temperature, the higher its value 
the more an object's orbit deviates from a circular orbit. 
 For low values of $v_\perp$, the deviation from $\Theta_{\rm LSR}$ 
gives information about the kinetic temperature.
As can be easily seen, most of the values
cluster around the LSR on a banana shaped region alongside the
\mbox{$v_\perp=0$~\kms} isovelocity line. This clustering means
that the majority of them is kinematically relatively cool. 
 A few stars are located further away, and in some cases
quite far away from the $v_\perp=0$~\kms\ contour. These are the
kinematically hot stars. Another reason that most of our stars have
quite low $v_\perp$ values is that most stars are near their
orbital turning points  (see the top of this subsection), 
i.e. their apo- or perigalacticon. In
these orbital phases the $\Phi$ component is minimised.

As the current $\Theta$ and $E_{\rm kin}$ are just snapshot
values, we also calculated the median of both quantities using the
whole orbit and plotted these in Fig.\,\ref{Etheta.fig} as well.
These data points lie nicely on a line pointing towards lower
$\Theta$ and lower $E_{\rm kin}$ away from the LSR.
There is a gap near $\Theta=110$ km\,s$^{-1}$. Stars located to the left
of this gap have rather hot orbits. Their considerable dispersion
in inclination is shown by the spread along the line (caused by
the $W$ velocity component). They represent the Halo population.
The stars at $\Theta>110$ \kms\ are the Disk stars. Clearly the
asymmetric drift of each star can be seen. The warmer an orbit is
kinematically (seen by increasing $v_\perp$), the lower its
orbital velocity is.

Considering $E_{\rm kin}$ and $\Theta_{\rm med}$ in Fig.\,\ref{Etheta.fig},
there is (similar to what we saw in Figs.\,\ref{eccnzehist.fig} and
\ref{nzeecctheta.fig}) a clear division
between the Halo and Thick Disk with the division
line here being at $\Theta=110$ km\,s$^{-1}$. Dividing Thick
and Thin Disk is not so straightforward.
Perhaps only the distribution of a sample complete
to the Galactic plane will lead to a separation of these
two kinematical populations.

{\it To conclude}, analysing the kinematics of the whole orbits
instead of just the current velocity gives further insight into the
kinematic behaviour of a group of stars. In our case the Halo and
Disk components of our sample are clearly discerned 
(Figs.\,\ref{eccnzehist.fig}, \ref{nzeecctheta.fig} \& \ref{Etheta.fig}).
Of the 114 stars, 16 (14\%) belong to the hitherto unknown Halo
group, the rest are Disk stars\footnote{A further 
candidate may be HE~2349$-$3135, which, 
while having $\Theta$=256 \kms\ and an orbit 
eccentricity of 0.49 (these values would classify it as a Disk star), 
travels to 12 kpc above the Galactic plane. 
This is rather typical for Halo stars. Therefore its status is unclear; 
we do not count it as a Halo star when stating the percentage of Halo stars, 
but do include it in our kinetically selected Halo sample 
which is also used for the scale height determination.}.
Thin and Thick Disk groups cannot really
be separated, at least not with the current sample, which is
perhaps lacking in Thin Disk stars. Possibly the kinematics of
both disk components are not really disjunct so that for a detailed
analysis statistically complete distributions are required.

\section{Determining a scale height for the stars using their orbits}
\label{scaleheight}

\begin{figure}
   \resizebox{\hsize}{!}{\epsfig{file=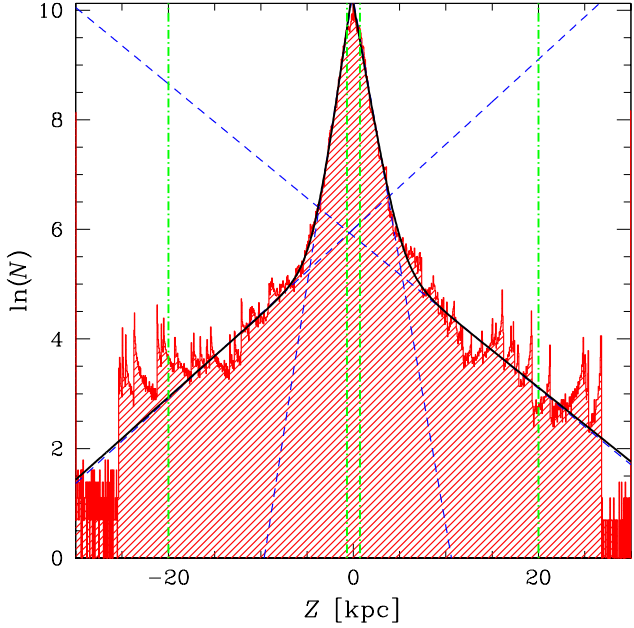}}
  \caption{  Histogram of the $z$-distance statistics of
   all the stars of the sample. The logarithmic distribution
   clearly shows the two components (see also Sect.\,\ref{scaleintro}).
    Fits using data within the dash-dotted verticals lead to the scale heights
   as given in Table\,\ref{sdB_results.tab}. The binsize is 50 pc.
               }
   \label{allhist.fig}
\end{figure}

\subsection{$z$-probability plot and scale height}
\label{scaleintro}
 We have derived the $z$-distribution\footnote{$z$ is the (positive) 
distance between the Galactic plane and a point (such as a star)
while $Z$ means the $Z$-coordinate of the point in the $XYZ,UVW$ system.
Technically we are determining the $Z$-probability distribution, 
because we are measuring the slope above and below the Galactic plane, 
i.e. for $Z<0$ kpc and $Z>0$ kpc. Assuming the symmetry of the disk 
in $Z$ direction both slopes will have similar values and can therefore 
be averaged to one, namely the $z$-distribution.} 
of our sample of 114 stars using the orbits of the stars, 
as done before in \citet{B97} for 41 stars. 
The program used to calculate the orbits does so for a fixed time per step, 
in our case 1~Myr. Plotting a histogram of the orbit in $\varpi$ or $z$ 
leads to the $\varpi$ resp.\ $z$ probability distribution for the star, 
i.e. the probability to find the star at a certain $\varpi$ or $z$ distance.
Plotting the histogram for the whole sample 
(by adding up the individual histograms) 
leads to the probability distribution for the complete group. 
    From this the $z$-density gradient of the sample can be deduced.
One may now fit an exponential distribution and determine the scale height
using the relation
\begin{equation}
       \ln{N(z)}=\ln{N_0}-\frac{z}{z_0},
\end{equation}
with $N(z)$ being the number density at $z$, $N_0$ being the
density at $z=0$ kpc, and $z_0$ the scale height. The scale height
is the reciprocal value of the slope of the $\ln N$ distribution.
As we do not have a defined volume in which the stars are located
we are unable to determine absolute values for $N(z)$ and $N_0$.
What we can determine, however, are relative values of the form
$N_1(0)/N_2(0)$ in the case that there are two or more slopes.

The method is described in greater detail in \citet{B97}. The $z$-distribution
is depicted in logarithmic form in 
Fig.\,\ref{allhist.fig}. 

Fig.\,\ref{allhist.fig} clearly
shows that our distribution consists of
two components with different $z$-distributions, a central one with a
steep slope and an outer shallow distribution.
 Fitting linear equations to the various regions
leads us to scale heights of 1.0~($\pm 0.1$)~kpc for the central (steep) part 
and 7~($\pm 2$)~kpc for the outer (shallow) parts. 
For the fit of the steep slope we used a fitting interval of [0.7,4] and
[$-$4,$-$0.7] kpc and [7,17] resp. [$-$18,$-$7] kpc for the shallow 
slope\footnote{The intervals used for this component are slightly 
different because of a disturbing spike in the one direction, 
which would somewhat falsify the fit result.}. 
   The results for the upper ($Z>0$ kpc) and lower ($Z<0$ kpc)
 half of the distribution are:

for the central part:\\
\hspace*{15mm}$z_{0,+}$~=~1.04~kpc, $z_{0,-}$~=~0.93~kpc

for the outer parts:\\
\hspace*{15mm}$z_{0,+}$~=~7.2~kpc, $z_{0,-}$~=~6.5~kpc. \\
The main reason for the greater uncertainty for the outer part is
that this relies on only a small group of stars and therefore
suffers from small number statistics. On the whole the result for
the component with the steep slope is very similar to that of
\citet{B97} based on a sample only 1/3 of the size of the present.
  The other component was not discernible for \citet{B97}, 
because their sample had only few data points at $z>$~2~kpc.

For small $z$ the distribution is less well known.
Here selection effects come into play (see Sect.\,\ref{samplesel}). 

\subsection{Effects of Errors}
\label{scaleerrors}

 The effects of the errors of the input parameters, i.e.
distance, radial velocity, and proper motion 
on the derived scale height was analysed by \citet{B97}.
 They added errors to these quantities,
and then computed the scale heights anew.
 The most important effect may be 
a systematic error in the distances: the scale height would decrease 
if the distances were systematically too small, and vice versa. 
The effects of the other quantities are smaller, 
with the scale height being smallest
when no error is added. As we expect similar effects 
on our sample, we did not repeat the
error analysis but refer to \citet{B97}. 

\subsection{Scale height and galactocentric distance}
\label{scalewbins}
To see how the scale height results vary with 
galactocentric distance we cut the cumulative orbit
file of our whole sample in bins in galactocentric 
distance ($\varpi$) and determined 
the scale heights for each. 
In all bins (except the outermost bin $\varpi \ge 15$ kpc, 
where the inner, steep distribution is missing) 
both components were found; 
however the outer, shallow component was overwhelmed by the steep component 
and thus hard to identify in the bins near $\varpi \simeq 8.5$ kpc. 
The results for the scale height of the steep
 component is  only marginally smaller than 
the result derived from all data points. 
Several aspects have to be taken into account here 
such as the increase in scale height with $\varpi$ 
and the increased  number of solar like orbits
in the central bins, 
 two aspects which counteract each other because the 
latter orbits have a smaller vertical extent.
Therefore no significant trend with $\varpi$ 
can be seen in the steep scale height. 

The extended, high $z$ distribution is based on only a few stars. 
For this reason one does expect a large spread in the derived
scale heights of the subsamples. 
Apart from this there is a trend from small to large scale heights 
with $\varpi$. 
This is 
an effect of the diminishing gravitational potential with 
galactocentric distance. 

\subsection{Robustness of the scale heights, separating the different populations}
\label{scalerobust}
   In order to test the stability of the values of our scale heights we made
tests with different fit interval limits. 
Furthermore, to determine how the various components influence each other, 
we derived scale heights with various subsamples excluded.

Shifting the lower and upper border of the fit interval of the steep component 
leads to a variation of the result of about 100~pc.
 The choice of interval width used for fitting 
leads to an error of $\sim50$~pc, 
which means that the derived slope and hence scale height 
is relatively robust in this respect. 
The fit intervals used to determine the slope have a larger influence 
on the Halo component because of various peaks caused by individual stars.

The presence of the Halo component has some influence on 
the scale height obtained for the Thick Disk. 
Without it (selected using the results of Sect.\,\ref{kinorb}) 
the scale height tends to be 100 - 150~pc lower. 
For the complete disk sample without the 17 stars classified as Halo members, 
we find a scale height of the steep component of 0.84$\pm0.1$~kpc instead 
of the 0.98$\pm0.1$~kpc for the whole sample. 
For this reason the Halo component should be removed before calculating 
the scale height of the Thick Disk. 

 Removing the stars with the coolest, i.e. most solar-like, 
kinematics (thus excluding stars with $ecc<0.15$ and $nze<0.15$) 
leads to a slightly higher scale height of just over 0.9 kpc. 
Obviously the influence of Thin Disk stars is much lower. 
This becomes quite clear, when one considers that removing stars 
with a $nze$ lower than a certain value only changes the distribution 
in the middle. 
 Therefore, the selection effects described in Sect.\,\ref{samplesel} 
do not play a significant role for the determination of the scale height. 
If the sample were complete for the low $z$ stars, 
the central part would fill in and possibly result in a third perhaps 
still steeper component, representing the Thin Disk.

The end points of the fit are more significant for the Halo component 
than for the disk part. 
This is expected as spikes in the distribution caused by individual stars
are important $-$ this subsample contains only 17 stars. 
On the other hand the influence
of the disk on the Halo scale height is small.

Another approach is 
to fit a 2-component function to this. 
This double exponential function 
(which accounts for the overlap in the region where both components are
of similar strength) has the following form:
\begin{equation}
\ln{N(z)}=\ln({N_{{\rm D},0}\cdot e^{-z/z_{\rm D}}+N_{{\rm H},0}\cdot e^{-z/z_{\rm H}}})
\end{equation}
with the indices D,H referring to the {\bf D}isk and {\bf H}alo respectively.
The resulting fit of this equation to the data is shown 
in Fig.\,\ref{allhist.fig}.
Here we used fitting intervals of [0.7,17] kpc and [$-$18,$-$0.7] kpc 
(shown in Fig.\,\ref{allhist.fig}).
 Varying the fitting intervals lead to the following results:
Raising the lower limits causes the scaleheight to slowly rise. 
This trend was mainly caused by one side of the distribution, 
the other one did not change. The upper limit has the same difficulties
caused by the spikes as in the single exponential fits.

The resulting values for the scale heights are:\\
\hspace*{15mm} $z_{\rm D}$=0.93$\pm0.09$~kpc, \\
\hspace*{15mm} $z_{\rm H}$=7.0 $\pm0.5$~kpc.\\
This means that there is a difference of 0.09~kpc between the Disk scale 
heights derived using population separation and the 2-component fit. 
The difference between the two values for the Halo is as large as 0.5 kpc; 
this may be related to the effect we see for the Disk scale height, 
but as the values for the Halo are not very certain
and depend much on the fit intervals, 
we do not further discuss the differences of the Halo scale height values.
Reasons for this small discrepancy could lie in the fit, 
as the presence of the Halo component may have an influence on the derived 
value for the Disk, 
fitting
intervals could still play a role here.

To conclude, we consider the 2-component method the most trustworthy and 
finalise the value of the scale heights as:\\
\hspace*{15mm}0.93$\pm0.08$ kpc for the Disk (or 0.9$\pm0.1$ kpc),\\
\hspace*{15mm}7.0$\pm1.0$ kpc for the Halo component.\footnote{This error is an 
estimate, but certainly more appropriate than the values derived 
from the difference of the two fits (upper and lower half) given above.
The value for the Halo part varies more than the 0.5 kpc when fitted under different
fitting circumstances.}

\begin{table}
\caption[Compilation of our results for the scale heights and mid plane densities]
{Compilation  of our results for the scale heights ($z_0$) 
and mid-plane densities $N_0$ for Thick Disk and Halo. 
The mid-plane density of the Halo is presented as the density ratio 
of Halo and Thick Disk.}
\begin{tabular}{p{4.1cm}clc}
\hline
Method                  &   \multicolumn{1}{c}{Thick Disk}
                        &   \multicolumn{2}{c}{Halo} \\
                        &  $z_0$   &   $N_0$ & $z_0$ \\
                        &   [kpc]   &  [\%]   & [kpc] \\
\hline
separate linear equations & 0.98 & 1.3 & 6.8 \\
2-component fit           & 0.93 & 1.2 & 7.0 \\
adopted values$^a$        & 0.93 & 1.2 & 7.0 \\
\hline
\end{tabular}\\
$^a$ These values rely on the 2-component fit.
\label{sdB_results.tab}
\end{table}

\subsection{Midplane number ratio of the two components found}
\label{scalenumrat}

The steep distribution has a peak value of $\ln N_0$~=~10.3 
and the shallow part peaks at $\ln N_0$~=~6.5. 
Therefore the relative density of the shallow with respect
to the steep component is 1.25$\pm$0.25\%.

The zero point of the broad distribution has a large uncertainty 
because it relies on the data of relatively few objects. 
Therefore the ratio of densities is only an estimate.

\subsection{Discussion of results}
\label{scaleres}

Our scale height study resulted in finding two components,
one with a large and one with a moderate
scale height.
A compilation of the results with the different strategies described 
in the sections above,  is given in Table\,\ref{sdB_results.tab}.
The latter scale height is with $z_0~=$~0.9 kpc
very similar to that found for the sdB stars by \citet{B97}. 
It is also similar to the determinations of the scale height of the Thick Disk 
(see e.g. \citealt{O94a}, \citealt{Kerber2001}, \citealt{Chen2001} 
and discussion below).
 Thus the steep component in Fig.\,\ref{allhist.fig} can be
identified with the Galactic Thick Disk population.

The analysis of the kinematics (Sect.~\ref{kinorb})
showed that there are some stars with solar kinematics. 
These stand out as an extra double peak in the centre of the linear distribution -
also vaguely seen near the peak of the logarithmic histogram of Fig.\,\ref{allhist.fig}. 
Presumably these and perhaps some more belong to the Old Thin Disk.

The shallow, high $z$ component 
represents a drastically different population of stars. 
 It is rather hard to imagine a population with such a scale height to 
be disk-like if one considers the Galaxy's radius being of the order 15 kpc.
So we speculate that this component is actually a spheroid or an ellipsoid. 
To make significant quantitative statements about the shape of 
the distribution of this population more stars are needed 
than the few which are discussed here. 
 But this subsample clearly consists of members of the Halo.

The density ratio of the Halo to the Disk component extrapolated 
to the Galactic plane is 1.2\% (see Table\,\ref{sdB_results.tab}.
 This value is of course quite uncertain. 
\citet{Kerber2001} found a density ratio of 0.2\% for Halo and Thin Disk. 
\citet{Chen2001} found a local relative density of the Halo against 
the Thin Disk of 0.125\%, in total values between 0.05\% and 0.4\% are found.
Our higher values for the Halo to Disk density is probably due to the
fact that our Disk stars are rather members of the Thick than the Thin Disk 
(or a mixture of both). 
The literature values for the Thick/Thin Disk density ratio 
range from about 5\% to 10\%. 
Assuming now a mean Thick/Thin Disk relative density of 7.5\%,
and a Halo/Thin Disk ration of 0.15\% (see above) 
the Halo/Thick Disk ratio becomes approximately 2.0\%.
This is - given the large spread in the literature values -
rather similar to the value we derived from the sdB stars.
Therefore the production rate of sdB stars seems to be of equal magnitude in both 
populations. Given the inherent different nature
 of stars in both Galactic components, such 
as metallicity, we conclude that these do not play a large role in the formation process
of sdB stars. For the Thin Disk we can unfortunately not make any definite statement
about the production rate of sdB stars in respect to the other populations. 

\section{Discussion: kinematics and the population membership of sdB stars}
\label{discussion}
As shown in Sections \ref{kinorb} and \ref{scaleheight} the sdB stars of our sample
belong to different populations. In this Section we want to discuss the different groups and
implications for the evolutionary processes that lead to the formation of sdB stars and
questions of Galactic structure. 

\subsection{Kinematical biases}
\label{kinbias}
Apart from the selection effects discussed in Sect.\,\ref{samplesel} which are mainly caused by
the sample composition, there are other effects resulting from the kinematics and orbital
morphologies themselves. These therefore not only occur in studies of specific object types but
in any kinematic study such as the one presented here.

Stars with different vectors of motion (in respect to the Sun)
have different probabilities of venturing into the vicinity of the
Sun (defined by the observable range of this study). An object
with a motion differing a lot from the solar motion will only stay
in the solar vicinity for a short while; a star with a solar like
orbit will remain there for a very long time. One the other hand
an object of the ``fast moving'' type (i.e. relative to the Sun)
has a greater probability of moving into the solar vicinity,
because a ``solar like'' object, once it is outside the solar
range, will stay outside for a long time. Presumably these two
effects\footnote{This is in principle the old problem of the two
clocks: Which one is better, a clock being slow (or fast) one
minute per hour or one that is completely broken?} compensate each
other or nearly do so; therefore we do not expect significant
skewness in our analyses caused by this. However stars with
extreme kinematics might very well be affected. This especially
applies to stars on retrograde orbits or in principle to stars
orbiting significantly faster than the Sun. 

A similar phenomenon could occur when looking at the motion perpendicular to the Galactic
plane. Here stars on polar orbits are presumably underrepresented in respect to stars travelling
on an orbit with a smaller inclination angle. Therefore spherical populations such as the
Halo may appear flattened somewhat when examining the distributions of their stars using the
kinematics. The effect on disk-like populations such as the Disks should be of much smaller
magnitude. Therefore we do not take it into account in the present study.

We do not quantitatively analyse probable effects caused by the selection effects
discussed in this subsection, because the focus of this study was to determine quantities
like the scale height of the Disk, which is not as much affected by such a bias, and to
find a possible Halo population. The number of Halo stars is too small to give more than rather
crude values for the scale height and other quantities.
However we do point out that in the future (when yet larger samples
give us the opportunity of a more detailed analysis of the Halo population)
it is certainly worthwhile if not essential
to analyse and quantify the biases introduced by such phenomena.

\subsection{The Disk}
\label{discdisk}

The vast majority of the sdB stars was found to belong to the Galactic Disk.
While most stars have orbits of moderate eccentricity and reach normalised $z$-heights ($nze$)
of around 0.2,
there are quite a few which have near solar kinematics, and are therefore more likely
to be associated with the Thin Disk. Unfortunately, the sample composition prevents us from
separating the two populations unambiguously if these are separate at all
(see Sect.\,\ref{samplesel}).
 The histogram in Fig.\,\ref{velhist.fig}
shows a Thick Disk-like distribution with a Thin Disk peak. This implies that
the stars of our sample come from both Disk components. Because of our result for the scale height
and the velocity dispersions, we conclude that the majority of the sample belongs to the Thick Disk.

On the whole most other studies arrive at values
 similar to what we found for the scale height and kinematics
of the Thick Disk. There is however 
some disagreement among these
studies concerning the scale height and hence the local relative density. \citet{Kerber2001}
derive a scale height for the Thick Disk of between 0.8 and 1.2~kpc. \citet{Reyle2001} 
 favour a value
of 0.8~kpc, \citet{Chen2001} 
arrive at 0.58 - 0.75~kpc but with a much
higher local density of 6.5 - 13\% of the density of the Thick Disk.
In contrast to those results, \citet{Reidmaj1993}\ 
 find a scale height of 1.4~kpc, in this case
the local density being not large, $\sim$2\%. As can be seen the values differ by at least
a factor of 2. Our value of 1~kpc is only a little above the values obtained in most studies.

The velocities and velocity dispersions are under much less dispute than the scale height.
Most studies arrive at values near those of \citet{O94b}, 
 i.e. $\sim50$~km\,s$^{-1}$ for each $UVW$ component and a value 
of 175~km\,s$^{-1}$ for the mean orbital velocity of the Thick Disk.
 Our larger value for $\bar{\Theta}$ and $\sigma_{\Theta}$
 is related to the Thin Disk
and Halo contamination. When we only consider stars with $ecc\le$~0.55 thus excluding
most of the Halo objects, we obtain values for $\sigma_{\Theta}$ of
about 50 to 60~km\,s$^{-1}$ (see Table\,\ref{vel.tab}).

\subsection{The Halo}
\label{discHalo}

As described in Sect.\,\ref{kinorb} and \ref{scaleheight}, the Halo
component stars can relatively easily be separated from the Disk population. 
 Of the complete sample of 114 stars,  16 (see Sect.\,\ref{orbvelana}) 
or $\sim$15\% form our Halo sample.
It is approximately divided in half, one part having small orbital velocities, 
i.e. having orbits similar to those of the HB stars of \citet{AB2000}, 
and the other, the high velocity component, 
having $\Theta$ velocities far larger than that of the LSR.
As can be seen in Fig.\,\ref{Etheta.fig} there is no 
or only a small difference between the kinematics of the two groups.
They represent orbits in different phases $-$ the low velocity ones 
are currently near their apogalacticon 
and the others are close to the perigalacticon where $\Theta$ is highest.
Therefore the orbits of both groups of the stars of our sample cover different
radial spaces in the Milky Way.
That they form a bimodal velocity distribution rather than a continuum 
in our sample  may be a consequence of the selection effects discussed 
in Sect.\,\ref{kinbias} and the small number statistics.

However some of the orbits of the high velocity group extend very far out,
further out than what
is generally considered to be the limits of the Galaxy.
Their orbits are rather transient than
bound orbits. So one might speculate that the
history of these stars is somewhat different than
that of the low velocity Halo stars.

It has long been known that stellar groups exist
at distances large compared with what is regarded as
the normal extent of the Milky Way. \citet{Harris} lists 9 globular clusters with a current
galactocentric distance of more than 30~kpc, the most remote objects being
at 100~kpc or more.
 \citet{Din991} have 6 objects in their sample of 38
globulars with kinematic data which go beyond 30 kpc. With one object (Pal 3)
in common that makes 14 globular clusters of the 147 listed in \citet{Harris},
venturing that far. However not only globular clusters but also field stars have been
found that far from the Galactic centre. \citet{Vivas2001} found $\sim150$ RR Lyrae
stars at distances of about 50~kpc and \citet{Yanny2000} found a large number
of HBA stars forming a group or a stream at a similar distance.
  
One of the Sun's neighbours, Barnards star\footnote{orbit calculated from Hipparcos data} features an orbit very similar to that of the stars of 
our ``high velocity Halo''. It travels to an apogalactic distance of 36~kpc, and has a $\Theta$ of 367~\kms. 
 So clearly this remote spatial regime is not unpopulated.

Our group of Halo and high velocity Halo stars is too small to make definite 
statements about the origin and behaviour of distant stellar groups as a whole.
However we speculate that two of our stars might have a
common origin because their trajectories are quite similar and extreme.
Analysing a larger sample might therefore give insight into moving streams of Halo
stars which are being incorporated into our Galaxy (see e.g. \citealt{Helmi1999}).

\subsection{Aspects of the stellar evolution history of sdB stars}
\label{discevol}

  Apart from questions concerning the structure of the Galaxy, there are still
aspects concerning the evolution of stars to sdB stars
which have not yet been satisfactorily solved. 
This especially applies to their extreme mass loss leaving a He-core nearly
completely stripped of all hydrogen.
It has been suspected for a long time that a large part, or all, 
of the sdB stars are in fact products of binary evolution 
(see e.g. \citealt{Iben1987}),
 with their unusually thin H-shell being the result of mass transfer 
from the evolving primary (the star later turning into the sdB star). 
In fact many sdBs show a secondary component in their spectrum 
and also in their colour indices (see e.g. \citealt{Thejll1994,PV,Aznar2002}). 
As described in Sect.\,\ref{dataredspec} many, however by far not all
(or even the majority) of the  apparently single sdB stars
have radial velocity variations 
revealing an unseen companion, such as a low mass main sequence star or a white
dwarf.
This means that it is certainly not clear whether close binary evolution 
is the only way sdB stars are formed.

A kinematical study does not as such prove or disprove a theory
about stellar evolution. However our results show that sdBs occur
in all locally observable older populations rather than in
one alone. This means that it is unlikely that there are factors
such as metallicity in play. Furthermore sdBs form from stars of a
quite significant spread in mass and hence in age,
 and the formation times of Halo, Thick and Thin Disk 
are of ages differing by several Gyr. Therefore our results
provide at least some support for the binary scenario.

\citet{Dorman1} have calculated models of horizontal branches for 
various metallicities ranging from very metal-poor to supersolar metallicity. 
These show a thinning out of the occupation of the HB in the middle, 
increasing with metallicity.
 \citet{DCruz1996} have further enhanced these models, and tried to find 
an explanation for the extreme mass loss required to make HBB and sdB stars.
The models were calculated for masses of around 1\,M$_\odot$ and somewhat less
massive for the metal-poor stars, 
taking into account that Halo stars are generally assumed to be
older than the more metal-rich stars of the Thick Disk 
and even more than those with solar metallicity.

Our present work shows that sdB stars can be found in all populations. 
The earlier work \citep{AB2000}, analysing the kinematics of all types 
of HB-stars in the temperature range between the RR-Lyrae and the sdB regimes, 
came to the result that very few if any HBA stars are Disk stars 
(with relatively high metallicities) 
and that only few RR-Lyraes with near solar metallicities 
and disk-like kinematics exist. This is well in line with the models showing
the deficiency in stars of the middle temperature range of 
the horizontal branch also in the data.

The bottom line is that our results actually fit to both
scenarios, the binary evolution as well as the RGB-peel-off mechanism of
\citet{DCruz1996}. Therefore we cannot prove or disprove one or
the other, even both could be in play. A still larger low
$z$-sample, which does not suffer from selection effects against
low $z_{\rm max}$ stars (see Sect.\,\ref{samplesel}) could help
answering this question. If the binary scenario is the dominating
process leading to the forming of sdB stars, then the ratio of sdB
stars belonging to the Halo, Thick and Thin Disk should be similar
to that of other evolved stars. If, however, the
\citet{DCruz1996} scheme holds true, then sdB stars in the Disk
should be a little more numerous than expected. A really large and
complete sample is required to find such subtle differences.

\section{Conclusions and outlook}
\label{conclusions}

In this study we demonstrated that there are not only sdB stars belonging
to the Disk, but that also a Halo Population exists.
The scale height of 0.9 kpc we derived for the Thick Disk 
is close to the middle of the range of 
values found in the literature. The trend in kinematics of stars 
along the horizontal branch described earlier \citep{AB2000} 
is still obvious in spite of the fact that 
we found a number of Halo sdB stars in this current study.
This hints to somewhat different routes in development from the RGB to the HB 
for stars which become HBA and stars become sdB. 

We have shown that sdB stars are a good tracer for the older 
populations. They are relatively numerous in the Thick Disk and
present in the Halo. Some more problematic issues remain. 
Due to the binary nature (with the secondary component unseen) 
of a significant fraction of sdB stars, individual objects may have 
a measured radial velocity which is far off the actual systemic value. 
 Another point is that we only had one first epoch position at our disposal. 
Therefore systemic radial velocities and new (possibly satellite based) 
proper motions should be obtained in the future.
Apart from this, sdB stars at low Galactic latitudes should be included 
in a similar study, to investigate the Thin Disk population. 
 Many more (even close by) sdB stars are not included, 
mostly because no measured radial velocity is available 
(although some of these have Tycho 2 or even Hipparcos proper motions).
Analyses similar to this one could also be carried out for 
other HB-like objects, 
whose kinematics have not yet been studied in great detail. 
Studies concerning the kinetic
behaviour of white dwarfs are underway (see, e.g. \citealt{Pauli2002}).
Once satellite missions such as GAIA are completed, 
we will have a much larger sample with much more accurate data. 
This would enable us to confine principal morphological parameters 
very closely, and give us access to substructure in all populations discussed.

\begin{acknowledgements}
      Part of this work was supported by the German
      \emph{Deut\-sche For\-schungs\-ge\-mein\-schaft, DFG\/} 
      (Bo~779/21 and He~1354/30-1).
      MA is supported by DLR grant No. 50QD 0102.
      We thank Tom Marsh (Southampton) for kindly supplying us with 
      his new systemic radial velocities. 
MA whishes to thank those contributing
      to the data collection in this long range project, 
namely Oliver Cordes, Andrea Dieball, J\"org
      Sanner, Ralf Vanscheidt, Yolanda Aguilar-Sanchez, Ralf Kohley,
      Armin Theissen, Sabine Moehler, Michael Lemke and Patrick Francois, 
and the assistants supporting
      observations at Calar Alto and La Silla observatories.
      Furthermore we wish to thank Michael Geffert, Uli Heber, Ralf Napiwotzki
      and Michael Odenkirchen for many fruitful discussions, the kinematic and astrometric software
      or spectral models they readily supplied to us. 
      With pleasure we made extensive use of
      the SIMBAD archive at CDS.
\end{acknowledgements}

\bibliographystyle{aa}
\bibliography{h4542.bib}

\end{document}

%% file: h4542t1a.tab
HE 0000$-$2355 & 00 03 22.06  & $-$23 38 58.0  & 13.29 & $-$0.25 & $-$0.13 & 0.019 & 0.76 &    +9.2 & $-$14.4 &   $-$64 &   \\
HE 0001$-$2443 & 00 04 31.00  & $-$24 26 21.2  & 13.88 & $-$0.26 & $-$0.14 & 0.019 & 0.82 &    +7.0 & $-$29.0 &     +20 &   \\
HE 0004$-$2737 & 00 06 46.26  & $-$27 20 53.4  & 13.97 & $-$0.28 & $-$0.16 & 0.019 & 0.71 &   +20.1 &  $-$7.6 &     +27 &   \\
PG 0004+133    & 00 07 33.77  &   +13 35 57.6  & 13.06 &         &         &       & 1.41 &    +3.0 & $-$25.0 &   $-$37 & B+97\\ 
HE 0021$-$2326 & 00 23 59.33  & $-$23 09 53.9  & 15.94 & $-$0.06 &   +0.01 & 0.018 & 2.72 &    +6.5 &  $-$1.3 &   $-$67 &   \\
HE 0031$-$2724 & 00 33 53.89  & $-$27 08 24.1  & 14.23 & $-$0.29 & $-$0.12 & 0.015 & 0.93 &  $-$2.7 &    +8.6 &   $-$12 &   \\
SB 290         & 00 42 58.31  & $-$38 07 37.3  & 10.40 &         &         & 0.013 & 0.26 &   +43.8 &  $-$7.0 &   $-$62 & HIP, AB00 \\ 
HD 4539        & 00 47 29.22  &   +09 58 55.7  & 10.31 &         &         &       & 0.17 &    +4.1 &   +24.0 &      +3 & HIP \\
PG 0039+049    & 00 42 06.11  &   +05 09 23.4  & 12.88 &         &         &       & 1.05 &    +7.5 & $-$12.0 &     +87 & B+97 \\ 
HE 0049$-$2928 & 00 51 57.74  & $-$29 12 07.5  & 15.78 & $-$0.22 & $-$0.12 & 0.014 & 2.20 &   +15.4 & $-$17.5 &   $-$15 &  \\
HE 0049$-$3059 & 00 51 37.70  & $-$30 42 56.2  & 14.41 & $-$0.24 & $-$0.13 & 0.017 & 1.22 &   +16.9 &  $-$2.1 &     +52 &  \\
SB 410         & 01 01 17.57  & $-$33 42 45.4  & 12.63 &         &         & 0.020 & 0.54 & $-$11.0 & $-$12.9 &   $-$56 & HIP   \\
Feige 11       & 01 04 21.67  &   +04 13 37.3  & 12.06 &         &         &       & 0.45 &   +12.2 & $-$40.0 &      +8 & B+97, M+99\\
SB 459         & 01 08 26.77  & $-$32 43 11.6  & 12.30 &         &         & 0.023 & 0.44 &  $-$7.7 &    +1.7 &   $-$24 & HIP \\
HE 0123$-$2808 & 01 25 33.35  & $-$27 53 04.7  & 16.09 & $-$0.24 & $-$0.07 & 0.016 & 2.61 &    +6.0 &  $-$3.0 &     +35 &  \\
HE 0127$-$4325 & 01 29 11.44  & $-$43 10 27.8  & 14.60 & $-$0.23 & $-$0.13 & 0.018 & 1.71 &    +8.3 &    +7.9 &     +16 &   \\
PG 0133+114    & 01 36 26.26  &   +11 39 31.0  & 12.28 &         &         &       & 0.77 &   +20.7 & $-$34.4 &      +6 &  M+02\\
PHL 1079       & 01 38 26.93  &   +03 39 38.0  & 13.28 &         &         &       & 0.81 &   +11.1 & $-$17.9 &       0 & B+97\\
HE 0136$-$2758 & 01 39 14.46  & $-$27 43 21.8  & 16.17 & $-$0.23 & $-$0.14 & 0.021 & 2.20 &   +15.6 & $-$22.9 &  $-$159 &  \\
SB 707         & 01 43 48.55  & $-$24 05 10.2  & 11.77 &         &         & 0.015 & 0.26 &   +88.4 & $-$46.2 &     +30 & HIP \\
PG 0142+148    & 01 45 39.57  &   +15 04 41.5  & 13.73 &         &         &       & 1.17 & $-$17.4 &  $-$0.4 &  $-$131 & B+97\\
SB 744         & 01 48 44.04  & $-$26 36 12.8  & 12.32 &         &         & 0.016 & 0.46 &   +90.6 & $-$47.0 &     +27 & HIP, KHD, UB\\
HE 0151$-$3919 & 01 53 11.20  & $-$39 04 18.0  & 14.31 & $-$0.22 & $-$0.10 & 0.014 & 2.11 &  $-$7.4 & $-$43.0 &  $-$176 &   \\
PG 0212+148    & 02 15 11.08  &   +15 00 04.6  & 14.48 &         &         &       & 1.75 &  $-$3.8 &  $-$9.2 &     +50 & B+97  \\
PG 0212+143    & 02 15 41.60  &   +14 29 18.0  & 14.58 &         &         &       & 1.85 &   +11.2 &  $-$1.4 &     +77 & B+97 \\
HE 0218$-$3437 & 02 20 59.75  & $-$34 23 35.2  & 13.39 & $-$0.25 & $-$0.12 & 0.018 & 0.81 &    +2.0 &  $-$2.8 &     +38 &  \\
HE 0218$-$4447 & 02 20 24.43  & $-$44 33 28.5  & 12.89 & $-$0.28 & $-$0.14 & 0.016 & 0.64 &   +47.9 &    +1.7 &   $-$15 &  \\
HE 0221$-$3250 & 02 23 58.15  & $-$32 36 32.6  & 14.70 & $-$0.24 & $-$0.10 & 0.016 & 1.60 &   +23.4 & $-$34.0 &   $-$73 &  \\
HE 0230$-$4323 & 02 32 54.68  & $-$43 10 27.8  & 13.78 & $-$0.22 & $-$0.11 & 0.023 & 1.03 & $-$11.7 &  $-$9.2 &  $-$102 &  \\
HE 0231$-$3441 & 02 34 00.25  & $-$34 27 54.9  & 14.83 & $-$0.24 & $-$0.13 & 0.022 & 1.02 &    +5.0 &   +20.3 &   $-$62 &  \\
PG 0242+132    & 02 45 38.86  &   +13 26 02.4  & 13.22 &         &         &       & 1.39 &   +17.2 &  $-$9.7 &     +11 & B+97 \\
HE 0258$-$2158 & 03 00 17.80  & $-$21 46 31.2  & 14.65 & $-$0.22 & $-$0.11 & 0.021 & 1.78 &  $-$0.8 &    +2.2 &     +72 &  \\
HE 0307$-$4554 & 03 09 25.93  & $-$45 43 33.0  & 15.06 & $-$0.21 & $-$0.10 & 0.022 & 1.02 &   +23.7 &  $-$7.4 &   $-$31 &  \\
HE 0315$-$4244 & 03 17 47.12  & $-$42 33 41.2  & 16.92 & $-$0.22 & $-$0.09 & 0.016 & 5.21 &    +4.1 &  $-$4.5 &    +121 &  \\
HE 0324$-$3749 & 03 26 14.98  & $-$25 18 38.0  & 14.62 & $-$0.24 & $-$0.13 & 0.017 & 1.14 &  $-$0.8 & $-$10.8 &     +74 &  \\
HE 0340$-$3820 & 03 42 47.06  & $-$38 11 26.4  & 14.77 & $-$0.28 & $-$0.15 & 0.010 & 1.43 &    +4.2 &    +7.8 &   $-$66 &  \\
HE 0341$-$2449 & 03 43 36.35  & $-$24 39 47.0  & 14.89 & $-$0.28 & $-$0.16 & 0.011 & 1.13 &   +20.0 &  $-$5.0 &     +10 &  \\
PG 0342+026    & 03 45 34.58  &   +02 47 52.8  & 10.94 &         &         &       & 0.36 &    +8.6 & $-$28.9 &     +13 & TYC, PI   \\
HE 0343$-$4748 & 03 45 09.52  & $-$47 38 54.0  & 14.19 & $-$0.17 & $-$0.07 & 0.006 & 2.12 &    +5.8 &  $-$2.7 &     +44 &  \\
HE 0351$-$3536 & 03 53 51.18  & $-$35 27 35.3  & 14.11 & $-$0.23 & $-$0.11 & 0.009 & 0.86 &  $-$0.6 &  $-$8.2 &     +64 &  \\
HE 0405$-$1719 & 04 07 27.54  & $-$17 11 15.5  & 14.00 & $-$0.28 & $-$0.14 & 0.025 & 0.91 &  $-$2.6 &   +15.8 &     +70 &  \\
HE 0405$-$3839 & 04 07 02.85  & $-$38 51 46.1  & 14.39 & $-$0.26 & $-$0.12 & 0.007 & 1.47 &    +5.2 &    +2.8 &   $-$44 & \\
HE 0407$-$1956 & 04 10 11.14  & $-$19 48 53.6  & 13.61 &         &         & 0.030 & 0.96 &   +13.9 &   +41.7 &    $-$6 & B+92 \\
HE 0410$-$4901 & 04 11 30.17  & $-$48 53 48.0  & 14.51 & $-$0.21 & $-$0.06 & 0.017 & 1.71 &    +5.4 &   +10.7 &   $-$27 &  \\
HE 0419$-$2538 & 04 22 04.17  & $-$25 31 00.5  & 13.67 & $-$0.23 & $-$0.12 & 0.044 & 1.29 &    +4.1 &  $-$8.2 &   $-$29 & \\
HE 0429$-$2448 & 04 31 28.29  & $-$24 41 56.8  & 15.30 & $-$0.24 & $-$0.12 & 0.048 & 1.18 &  $-$5.6 &  $-$9.0 &     +15 &  \\
HE 0442$-$1746 & 04 44 34.88  & $-$17 40 42.7  & 15.15 & $-$0.27 & $-$0.15 & 0.043 & 1.26 &  $-$5.3 &   +11.5 &     +18 &  \\
HE 0444$-$4945 & 04 46 14.09  & $-$49 40 10.8  & 15.11 & $-$0.26 & $-$0.12 & 0.011 & 1.53 &    +9.4 &   +23.9 &     +75 &  \\
HE 0447$-$3654 & 04 49 15.62  & $-$36 49 28.7  & 14.55 & $-$0.25 & $-$0.12 & 0.011 & 1.12 &   +16.5 & $-$20.1 &    +128 &  \\
HE 0452$-$3654 & 04 53 52.66  & $-$36 49 15.2  & 13.86 & $-$0.27 & $-$0.13 & 0.010 & 0.86 &    +7.4 &    +9.8 &   $-$62 &  \\
HE 0500$-$3518 & 05 02 31.67  & $-$35 14 19.4  & 15.04 & $-$0.23 & $-$0.11 & 0.013 & 1.29 &    +8.3 &   +15.8 &     +19 &  \\
HE 0504$-$2041 & 05 06 39.67  & $-$20 37 38.4  & 14.96 & $-$0.25 & $-$0.11 & 0.032 & 1.44 &    +7.1 &    +2.3 &   $-$11 &  \\
HE 0505$-$2228 & 05 07 47.46  & $-$22 24 27.6  & 15.55 & $-$0.26 & $-$0.09 & 0.034 & 1.19 &   +16.6 &    +1.1 &   $-$12 &  \\
HE 0505$-$3833 & 05 06 58.85  & $-$38 29 15.5  & 14.18 & $-$0.24 & $-$0.12 & 0.024 & 0.93 &    +3.1 &    +6.5 &     +69 &  \\
HE 0510$-$4023 & 05 12 18.20  & $-$40 19 34.5  & 14.84 & $-$0.22 & $-$0.08 & 0.048 & 1.51 &    +3.3 &  $-$6.2 &     +49 &  \\
HE 0516$-$2311 & 05 18 06.98  & $-$23 08 45.2  & 15.60 &         &         & 0.037 & 2.27 & $-$17.9 &   +17.8 &   $-$12 & HE \\
HE 0521$-$3914 & 05 23 25.48  & $-$39 11 54.3  & 15.55 &         &         & 0.026 & 1.82 & $-$15.0 &   +11.2 &   $-$87 & K+99 \\
HE 0523$-$1831 & 05 25 31.32  & $-$18 29 08.6  & 14.31 & $-$0.28 & $-$0.13 & 0.058 & 1.78 &    +4.8 & $-$10.1 &    $-$6 & \\
HE 0532$-$4503 & 05 33 40.50  & $-$45 01 35.3  & 16.06 & $-$0.22 & $-$0.11 & 0.037 & 2.56 &    +8.2 &  $-$2.7 &  $-$166 & \\

%% file: h4542t1b.tab
HE 0539$-$4246 & 05 41 06.69  & $-$42 45 31.9  & 14.80 & $-$0.20 & $-$0.11 & 0.045 & 1.28 &  $-$0.2 &    +3.7 &     +71 &   \\
PG 0856+121    & 08 59 02.72  &   +11 56 24.7  & 13.47 &         &         &       & 0.99 & $-$19.4 & $-$19.8 &     +85 & B+97 \\
PG 0907+123    & 09 10 07.60  &   +12 08 26.1  & 13.94 &         &         &       & 1.52 &    +6.4 &  $-$2.6 &     +85 & B+97, M+02\\
PG 0918+029    & 09 21 28.23  &   +02 46 02.2  & 13.33 &         &         &       & 1.04 & $-$28.5 & $-$20.0 &    +104 & B+97, TM \\
PG 0919+273    & 09 22 39.83  &   +27 02 26.2  & 12.76 &         &         &       & 0.35 &   +22.9 & $-$19.8 &   $-$65 & B+97 \\
PG 1101+249    & 11 04 31.73  &   +24 39 44.8  & 12.79 &         &         &       & 0.39 & $-$30.3 &   +16.0 &   $-$48 & B+97 \\
PG 1114+073    & 11 16 49.67  &   +06 59 30.8  & 13.06 &         &         &       & 0.45 & $-$12.3 & $-$14.4 &      +9 & B+97, TM \\
PG 1232$-$136  & 12 35 18.92  & $-$13 55 09.3  & 13.34 &         &         &       & 0.57 & $-$46.4 &  $-$1.7 &     +55 & B+97 \\
PG 1233+427    & 12 35 51.64  &   +42 22 42.6  & 12.04 &         &         &       & 0.32 &    +3.6 & $-$18.1 &     +61 & B+97 \\
Feige 66       & 12 37 23.52  &   +25 03 59.9  & 10.60 &         &         &       & 0.18 &    +2.7 & $-$26.7 &      +1 & HIP, AB00\\
PG 1256+278    & 12 59 21.27  &   +27 34 05.2  & 14.25 &         &         &       & 0.78 & $-$24.6 &    +3.5 &     +64 & B+97 \\
PG 1343$-$101  & 13 46 08.07  & $-$10 26 48.2  & 13.76 &         &         &       & 0.72 & $-$28.0 &  $-$3.7 &     +49 & B+97 \\
HD 127493      & 14 32 21.49  & $-$22 39 25.6  & 10.04 &         &         &       & 0.12 & $-$32.8 & $-$17.2 &     +13 & HIP, AB00\\
PG 1432+004    & 14 35 19.83  &   +00 13 48.0  & 12.76 &         &         &       & 0.76 &  $-$9.4 & $-$25.8 &      +1 & B+97, TM \\
PG 1433+239    & 14 35 20.36  &   +23 45 27.5  & 12.45 &         &         &       & 0.47 &  $-$3.5 & $-$18.5 &   $-$56 & B+97 \\
PG 1452+198    & 14 54 39.81  &   +19 37 00.9  & 12.48 &         &         &       & 0.81 &  $-$7.2 & $-$21.0 &     +51 & B+97 \\
PG 1519+640    & 15 20 31.32  &   +63 52 08.0  & 12.46 &         &         &       & 0.65 &   +28.1 &   +41.2 &      +2 & TYC, TM \\
PG 1619+522    & 16 20 38.74  &   +52 06 08.8  & 13.30 &         &         &       & 0.77 &  $-$3.6 &    +9.0 &   $-$52 & B+97, M+02\\
HD 149382      & 16 34 23.33  & $-$04 00 52.0  &  8.87 &         &         &       & 0.08 &  $-$6.0 &  $-$3.9 &      +3 & HIP, AB00\\
PG 1647+252    & 16 49 08.97  &   +25 10 05.7  & 14.07 &         &         &       & 0.71 &  $-$3.8 &   +12.3 &     +26 & B+97 \\
PG 1708+602    & 17 09 15.90  &   +60 10 10.8  & 13.48 &         &         &       & 1.79 & $-$14.9 &   +12.1 &    $-$8 & B+97 \\
PG 1710+490    & 17 12 18.74  &   +48 58 35.9  & 12.90 &         &         &       & 0.72 &   +10.8 &  $-$7.0 &   $-$54 & B+97, TM \\
PG 1716+426    & 17 18 03.54  &   +42 34 18.4  & 13.97 &         &         &       & 1.20 &    +7.1 & $-$21.8 &    $-$4 & G98, M+02\\
PG 1722+286    & 17 24 11.97  &   +28 35 26.9  & 13.24 &         &         &       & 0.87 &  $-$4.0 &   +10.0 &   $-$34 & B+97, TM \\
PG 1725+252    & 17 27 57.39  &   +25 08 35.7  & 12.89 &         &         &       & 0.66 & $-$17.7 &    +9.0 &   $-$60 & B+97, M+02\\
PG 1738+505    & 17 39 28.44  &   +50 29 25.1  & 13.15 &         &         &       & 0.97 &  $-$7.6 &    +9.0 &     +22 & B+97 \\
UV 1758+36     & 18 00 18.86  &   +36 28 56.3  & 11.37 &         &         &       & 0.20 & $-$28.2 &    +7.3 &       0 & HIP \\
HD 171858      & 18 37 56.68  & $-$23 11 35.2  &  9.85 &         &         &       & 0.16 & $-$15.0 & $-$24.7 &     +74 & HIP, M+02\\
HE 2135$-$3749 & 21 38 40.59  & $-$37 36 22.7  & 13.93 & $-$0.23 & $-$0.14 & 0.034 & 0.72 &    +5.3 &    +6.0 &  $-$156 &   \\
HD 205805      & 21 39 10.61  & $-$46 05 51.5  & 10.16 &         &         &       & 0.20 &   +76.4 &  $-$9.9 &   $-$57 & HIP, B+97\\
HE 2154$-$4143 & 21 58 01.98  & $-$41 28 49.7  & 15.17 & $-$0.25 & $-$0.12 & 0.019 & 1.49 &    +5.3 &    +5.7 &   $-$15 &   \\
HE 2155$-$1724 & 21 58 15.92  & $-$17 09 45.3  & 15.04 & $-$0.24 & $-$0.11 & 0.046 & 1.27 &   +11.6 &  $-$2.2 &   $-$27 &   \\
HE 2156$-$1732 & 21 59 30.14  & $-$17 18 21.6  & 14.45 & $-$0.22 & $-$0.08 & 0.041 & 1.39 &    +8.7 & $-$10.2 &   $-$77 &   \\
HE 2156$-$3927 & 21 59 35.50  & $-$39 13 14.8  & 15.06 & $-$0.19 & $-$0.06 & 0.019 & 1.40 &    +4.4 &   +22.0 &  $-$200 &   \\
HE 2201$-$2136 & 22 04 06.71  & $-$20 59 09.3  & 15.90 & $-$0.23 & $-$0.11 & 0.038 & 1.88 &    +9.2 &  $-$3.9 &   $-$23 &   \\
PG 2204+035    & 22 07 16.50  &   +03 42 19.8  & 14.25 &         &         &       & 1.18 &    +7.5 &  $-$6.0 &     +81 & B+97 \\
HE 2205$-$1952 & 22 08 41.30  & $-$19 37 39.4  & 14.59 & $-$0.28 & $-$0.14 & 0.023 & 1.01 &    +7.5 &  $-$8.8 &   $-$52 &   \\
HE 2213$-$4158 & 22 16 17.70  & $-$41 43 22.2  & 16.15 & $-$0.34 & $-$0.12 & 0.012 & 3.95 &    +5.6 &   +10.3 &    $-$4 &   \\
PG 2218+020    & 22 21 24.83  &   +02 16 18.6  & 14.17 &         &         &       & 1.15 &    +1.1 & $-$11.8 &     +21 & B+97 \\
HE 2222$-$3738 & 22 24 56.43  & $-$37 23 30.2  & 14.89 & $-$0.23 & $-$0.13 & 0.016 & 1.39 &   +32.0 &  $-$6.0 &  $-$134 &   \\
PG 2226+094    & 22 28 58.41  &   +09 37 21.8  & 14.08 &         &         &       & 1.13 &   +14.3 &    +0.4 &   $-$48 & B+97, TM \\
PG 2259+134    & 23 01 45.82  &   +13 38 37.5  & 14.48 &         &         &       & 1.38 &    +0.6 &  $-$9.6 &     +16 & B+97 \\
Feige 108$^b$  & 23 16 12.41  & $-$01 50 34.5  & 12.96 &         &         &       & 0.40 &  $-$7.8 & $-$16.5 &     +40 & B+97 \\
Feige 109$^b$  & 23 17 26.89  &   +07 52 04.9  & 13.76 &         &         &       & 1.13 &  $-$1.2 &    +8.1 &   $-$37 & B+97, TM \\
PG 2337+070$^b$& 23 40 04.83  &   +07 17 11.0  & 13.47 &         &         &       & 0.77 & $-$19.1 & $-$37.4 &   $-$27 & B+97 \\
HE 2337$-$2944 & 23 40 15.33  & $-$29 27 59.8  & 14.45 & $-$0.25 & $-$0.15 & 0.019 & 0.87 &   +14.3 &    +8.4 &   $-$26 &   \\
HE 2340$-$2806 & 23 42 41.39  & $-$27 50 01.6  & 15.01 & $-$0.16 & $-$0.05 & 0.017 & 1.46 &    +1.7 &   +11.3 &   $-$20 &   \\
SB 815         & 23 44 22.00  & $-$34 27 00.4  & 10.96 &         &         & 0.013 & 0.25 & $-$22.3 &  $-$7.2 &     +24 & HIP \\
HE 2343$-$2944 & 23 46 17.75  & $-$29 27 49.9  & 15.06 & $-$0.21 & $-$0.13 & 0.019 & 1.29 &    +5.5 & $-$15.2 &    $-$0 &   \\
HE 2349$-$3135 & 23 51 43.64  & $-$31 18 52.9  & 15.91 & $-$0.28 & $-$0.11 & 0.014 & 2.08 &  $-$6.6 &  $-$1.9 &    +180 &   \\
PG 2349+002$^b$& 23 51 53.26  &   +00 28 18.0  & 13.28 &         &         &       & 0.82 & $-$10.1 & $-$15.7 &   $-$84 & B+97 \\
SB 884         & 23 52 36.10  & $-$30 10 09.1  & 12.09 &         &         & 0.019 & 0.37 &   +30.4 & $-$13.2 &    $-$3 & HIP \\
HE 2355$-$3221 & 23 58 22.47  & $-$32 04 39.1  & 15.18 & $-$0.22 & $-$0.14 & 0.012 & 1.69 &  $-$0.1 & $-$15.5 &     +61 &  \\
PG 2358+107    & 00 01 06.73  &   +11 00 36.3  & 13.62 &         &         &       & 0.83 &  $-$3.0 & $-$14.0 &   $-$19 & B+97 \\
HE 2359$-$2844 & 00 01 38.46  & $-$28 27 42.8  & 16.62 &         &         & 0.019 & 1.53 &    +7.8 & $-$12.0 &  $-$129 & L+00 \\

%% file: h4542t2a.tab
HE 0000$-$2355 & $-$8.40 &   +0.11 & $-$0.75 &   $-$3 & +165  &   +61 &    +5  & +165 & +1382 &   8.42 &  5.46 &   1.63 &  0.20 &  0.21 \\ 
HE 0001$-$2443 & $-$8.39 &   +0.11 & $-$0.80 &   +42  & +127  & $-$23 & $-$40  & +127 & +1067 &   8.68 &  3.46 &   0.93 &  0.11 &  0.43 \\
HE 0004$-$2737 & $-$8.39 &   +0.06 & $-$0.70 &  $-$33 & +184  & $-$31 &   +34  & +183 & +1540 &   8.78 &  6.01 &   0.92 &  0.11 &  0.19 \\ 
   PG 0004+133 & $-$8.78 &   +0.90 & $-$1.05 &    +75 & +100  & $-$76 & $-$65  & +107 &  +941 &   9.43 &  2.92 &   3.48 &  0.41 &  0.53 \\ 
HE 0021$-$2326 & $-$8.35 &   +0.31 & $-$2.70 &  $-$58 & +171  &   +64 &   +64  & +169 & +1410 &   9.47 &  6.31 &   4.11 &  0.47 &  0.20 \\ 
HE 0031$-$2724 & $-$8.45 &   +0.03 & $-$0.93 &      0 & +273  &   +21 &    +1  & +273 & +2308 &  14.46 &  8.48 &   1.56 &  0.11 &  0.26 \\ 
   PG 0039+049 & $-$8.77 &   +0.49 & $-$0.89 &  $-$16 & +214  & $-$99 &   +28  & +213 & +1868 &  10.77 &  8.22 &   3.24 &  0.31 &  0.13 \\ 
 CD $-$38\,222 & $-$8.47 & $-$0.04 & $-$0.26 &  $-$40 & +208  &   +69 &   +39  & +208 & +1761 &   9.48 &  7.40 &   1.53 &  0.16 &  0.12 \\ 
       HD 4539 & $-$8.55 &   +0.09 & $-$0.14 &   $-$2 & +248  &   +17 &    +5  & +248 & +2120 &  11.09 &  8.55 &   0.30 &  0.03 &  0.13 \\ 
HE 0049$-$2928 & $-$8.46 & $-$0.07 & $-$2.20 &  $-$26 & $-$6  &   +30 &   +26  & $-$5 & $-$45 &   9.34 &  0.08 &   7.24 &  3.07 &  0.98 \\ 
HE 0049$-$3059 & $-$8.46 & $-$0.06 & $-$1.21 &  $-$63 & +169  & $-$43 &   +62  & +170 & +1435 &   9.54 &  5.27 &   1.65 &  0.17 &  0.29 \\ 
        SB 410 & $-$8.48 & $-$0.06 & $-$0.53 &    +50 & +230  &   +66 & $-$51  & +230 & +1950 &  11.42 &  7.66 &   1.78 &  0.16 &  0.20 \\ 
      Feige 11 & $-$8.65 &   +0.18 & $-$0.38 &    +25 & +162  & $-$42 & $-$22  & +163 & +1410 &   8.87 &  8.31 &   3.03 &  0.25 &  0.10 \\ 
        SB 459 & $-$8.50 & $-$0.05 & $-$0.44 &    +21 & +250  &   +31 & $-$23  & +250 & +2123 &  11.70 &  8.36 &   0.83 &  0.07 &  0.17 \\ 
HE 0123$-$2808 & $-$8.76 & $-$0.22 & $-$2.58 &  $-$31 & +158  & $-$17 &   +26  & +159 & +1392 &   9.23 &  5.55 &   2.66 &  0.30 &  0.25 \\ 
HE 0127$-$4325 & $-$8.41 & $-$0.51 & $-$1.63 &  $-$81 & +237  & $-$14 &   +67  & +242 & +2036 &  12.99 &  7.62 &   2.29 &  0.18 &  0.26 \\ 
   PG 0133+114 & $-$8.88 &   +0.32 & $-$0.59 &   $-$2 & +107  & $-$62 &    +6  & +106 &  +946 &   8.90 &  2.88 &   1.49 &  0.18 &  0.50 \\ 
      PHL 1079 & $-$8.86 &   +0.25 & $-$0.68 &     +9 & +159  & $-$19 &  $-$4  & +159 & +1410 &   8.88 &  5.13 &   0.79 &  0.09 &  0.27 \\ 
HE 0136$-$2758 & $-$8.82 & $-$0.25 & $-$2.16 &    +56 & $-$33 &  +194 & $-$55  &$-$35 &$-$307 &  11.24 &  4.05 &  11.27 &  3.60 &  0.47 \\ 
        SB 707 & $-$8.55 & $-$0.02 & $-$0.25 &  $-$43 & +121  &  $-$3 &   +43  & +121 & +1034 &   8.79 &  3.18 &   0.26 &  0.03 &  0.47 \\ 
   PG 0142+148 & $-$9.14 &   +0.50 & $-$0.84 &   +156 & +237  &   +80 &  +142  & +245 & +2245 &  20.76 &  6.91 &   3.90 &  0.19 &  0.50 \\ 
        SB 744 & $-$8.58 & $-$0.06 & $-$0.45 &  $-$78 &  +31  &   +23 &   +78  &  +31 &  +268 &   9.25 &  0.51 &   5.23 &  0.91 &  0.89 \\ 
HE 0151$-$3919 & $-$8.61 & $-$0.62 & $-$2.00 &   +362 &  +25  &  +239 &  +363  & $-$1 & $-$10 &  84.83 &  0.02 &  85.51 & 29.18 &  1.00 \\ 
   PG 0212+148 & $-$9.62 &   +0.61 & $-$1.20 &    +25 & +221  & $-$86 & $-$11  & +222 & +2143 &  11.84 &  9.69 &   3.42 &  0.30 &  0.10 \\ 
   PG 0212+143 & $-$9.68 &   +0.64 & $-$1.28 & $-$103 & +190  & $-$22 &  +115  & +182 & +1769 &  13.48 &  5.79 &   1.70 &  0.13 &  0.40 \\ 
HE 0218$-$3437 & $-$8.65 & $-$0.24 & $-$0.76 &     +6 & +211  & $-$24 & $-$12  & +211 & +1823 &   8.78 &  8.22 &   0.78 &  0.10 &  0.04 \\ 
HE 0218$-$4447 & $-$8.53 & $-$0.27 & $-$0.58 &  $-$93 & +149  &   +70 &   +88  & +152 & +1300 &   9.92 &  4.54 &   2.13 &  0.22 &  0.37 \\ 
HE 0221$-$3250 & $-$8.83 & $-$0.46 & $-$1.49 &    +89 & $-$41 &  +153 & $-$87  &$-$45 &$-$401 &  10.08 &  1.32 &   7.97 &  1.50 &  0.70 \\ 
HE 0230$-$4323 & $-$8.60 & $-$0.44 & $-$0.92 &    +91 & +289  &   +88 &  +106  & +284 & +2442 &  20.41 &  7.92 &   4.22 &  0.20 &  0.50 \\ 
HE 0231$-$3441 & $-$8.71 & $-$0.34 & $-$0.94 &  $-$65 & +306  &   +67 &   +53  & +308 & +2687 &  20.21 &  8.59 &   3.35 &  0.19 &  0.46 \\ 
   PG 0242+132 & $-$9.49 &   +0.34 & $-$0.91 &  $-$47 & +117  &    +9 &   +51  & +116 & +1099 &   9.90 &  3.76 &   0.98 &  0.11 &  0.45 \\ 
HE 0258$-$2158 & $-$9.27 & $-$0.44 & $-$1.55 &  $-$30 & +235  & $-$55 &   +19  & +236 & +2192 &  12.48 &  9.16 &   2.69 &  0.22 &  0.15 \\ 
HE 0307$-$4554 & $-$8.63 & $-$0.54 & $-$0.85 &  $-$19 & +156  &   +99 &    +9  & +157 & +1357 &   8.67 &  6.49 &   2.41 &  0.38 &  0.19 \\ 
HE 0315$-$4244 & $-$9.48 & $-$2.70 & $-$4.35 &    +26 &  +46  & $-$23 & $-$38  &  +37 &  +367 &  11.90 &  0.72 &   6.20 &  0.72 &  0.89 \\ 
HE 0324$-$3749 & $-$9.01 & $-$0.41 & $-$0.94 &    +24 & +173  & $-$62 & $-$31  & +172 & +1549 &   9.27 &  6.25 &   2.01 &  0.22 &  0.19 \\ 
HE 0340$-$3820 & $-$8.92 & $-$0.76 & $-$1.13 &  $-$28 & +279  &   +75 &    +4  & +280 & +2445 &  16.80 &  8.85 &   3.12 &  0.22 &  0.33 \\ 
HE 0341$-$2449 & $-$9.04 & $-$0.44 & $-$0.88 &  $-$22 & +143  &   +61 &   +15  & +144 & +1306 &   9.11 &  4.80 &   1.93 &  0.22 &  0.31 \\ 
   PG 0342+026 & $-$8.78 & $-$0.02 & $-$0.22 &    +16 & +188  & $-$15 & $-$16  & +188 & +1647 &   8.87 &  6.50 &   0.31 &  0.03 &  0.15 \\ 
HE 0343$-$4748 & $-$8.82 & $-$1.31 & $-$1.64 &     +3 & +158  &   +14 & $-$26  & +156 & +1390 &   9.18 &  5.17 &   1.69 &  0.19 &  0.28 \\ 
HE 0351$-$3536 & $-$8.80 & $-$0.46 & $-$0.66 &    +17 & +185  & $-$43 & $-$26  & +183 & +1616 &   9.02 &  6.50 &   1.19 &  0.13 &  0.16 \\ 
HE 0405$-$1719 & $-$9.06 & $-$0.34 & $-$0.63 &  $-$77 & +262  & $-$30 &   +67  & +264 & +2398 &  15.98 &  8.38 &   1.27 &  0.08 &  0.31 \\ 
HE 0405$-$3839 & $-$8.96 & $-$0.87 & $-$1.09 &   $-$6 & +247  &   +65 & $-$18  & +247 & +2220 &  12.97 &  8.76 &   2.06 &  0.20 &  0.19 \\ 
HE 0407$-$1956 & $-$9.06 & $-$0.39 & $-$0.66 & $-$146 & +318  &   +99 &  +132  & +324 & +2941 &  38.12 &  8.20 &   9.11 &  0.25 &  0.65 \\ 
HE 0410$-$4901 & $-$8.78 & $-$1.15 & $-$1.23 &  $-$81 & +256  &   +46 &   +47  & +264 & +2339 &  13.71 &  8.63 &   1.85 &  0.18 &  0.28 \\ 
HE 0419$-$2538 & $-$9.18 & $-$0.65 & $-$0.88 &    +58 & +205  &   +37 & $-$72  & +200 & +1842 &  11.36 &  6.82 &   1.34 &  0.12 &  0.25 \\ 
HE 0429$-$2448 & $-$9.15 & $-$0.61 & $-$0.77 &    +50 & +218  & $-$35 & $-$65  & +214 & +1963 &  11.45 &  7.58 &   1.38 &  0.12 &  0.20 \\ 
HE 0442$-$1746 & $-$9.33 & $-$0.60 & $-$0.74 &  $-$44 & +290  &  $-$6 &   +25  & +292 & +2728 &  19.23 &  9.29 &   1.35 &  0.07 &  0.35 \\ 
HE 0444$-$4945 & $-$8.77 & $-$1.14 & $-$0.99 & $-$185 & +186  &    +3 &  +160  & +208 & +1842 &  14.95 &  5.54 &   1.41 &  0.11 &  0.49 \\ 
HE 0447$-$3654 & $-$8.94 & $-$0.75 & $-$0.71 &    +38 &  +51  & $-$17 & $-$42  &  +47 &  +423 &   9.24 &  0.94 &   4.43 &  0.67 &  0.82 \\ 
HE 0452$-$3654 & $-$8.84 & $-$0.58 & $-$0.54 &   $-$7 & +275  &   +74 & $-$11  & +275 & +2438 &  16.46 &  8.84 &   2.80 &  0.17 &  0.30 \\ 
HE 0500$-$3518 & $-$9.05 & $-$0.88 & $-$0.77 &  $-$93 & +233  &   +51 &   +70  & +241 & +2193 &  13.82 &  8.17 &   2.07 &  0.15 &  0.26 \\ 
HE 0504$-$2041 & $-$9.41 & $-$0.81 & $-$0.76 &   $-$2 & +222  &   +57 & $-$17  & +221 & +2088 &  10.74 &  9.14 &   1.73 &  0.16 &  0.08 \\ 
HE 0505$-$2228 & $-$9.23 & $-$0.70 & $-$0.64 &   $-$1 & +190  &   +91 & $-$14  & +190 & +1758 &   9.54 &  8.01 &   2.71 &  0.29 &  0.09 \\ 
HE 0505$-$3833 & $-$8.85 & $-$0.66 & $-$0.55 &  $-$44 & +189  & $-$19 &   +30  & +191 & +1698 &   9.23 &  6.74 &   0.65 &  0.07 &  0.16 \\ 
HE 0510$-$4023 & $-$9.02 & $-$1.11 & $-$0.87 &    +32 & +170  &  $-$6 & $-$52  & +165 & +1497 &   9.71 &  5.34 &   0.95 &  0.10 &  0.29 \\ 
HE 0516$-$2311 & $-$9.88 & $-$1.39 & $-$1.14 & $-$114 & +457  & $-$88 &   +49  & +468 & +4672 & 161.35 &  9.94 &  26.79 &  0.17 &  0.88 \\ 
HE 0521$-$3914 & $-$9.17 & $-$1.37 & $-$1.00 &  $-$37 & +407  & $-$37 & $-$24  & +408 & +3781 &  64.93 &  9.31 &   7.35 &  0.11 &  0.75 \\ 

%% file: h4542t2b.tab
HE 0523$-$1831 & $-$9.70 & $-$1.04 & $-$0.81 &    +74 & +166  &   +16 & $-$91  & +157 & +1531 &  11.31 &  5.28 &   0.88 &  0.08 &  0.40 \\ 
HE 0532$-$4503 & $-$9.20 & $-$2.05 & $-$1.36 &    +83 & +305  &  +176 &  +147  & +280 & +2641 &  41.28 &  7.79 &  18.28 &  0.47 &  0.68 \\ 
HE 0539$-$4246 & $-$8.90 & $-$1.02 & $-$0.65 &  $-$34 & +185  & $-$26 &   +12  & +187 & +1678 &   9.05 &  6.77 &   0.82 &  0.09 &  0.14 \\ 
   PG 0856+121 & $-$9.16 & $-$0.49 &   +0.55 &  $-$74 & +116  & $-$46 &   +68  & +120 & +1099 &   9.80 &  3.43 &   1.33 &  0.14 &  0.48 \\ 
   PG 0907+123 & $-$9.47 & $-$0.75 &   +0.90 &   $-$6 & +177  &   +86 &  $-$8  & +176 & +1676 &   9.53 &  7.44 &   2.81 &  0.31 &  0.12 \\ 
   PG 0918+029 & $-$9.06 & $-$0.65 &   +0.59 & $-$101 &  +89  & $-$75 &   +94  &  +96 &  +874 &   9.46 &  6.88 &   4.77 &  0.28 &  0.58 \\ 
   PG 0919+273 & $-$8.74 & $-$0.09 &   +0.24 &    +87 & +221  & $-$16 & $-$89  & +220 & +1920 &  10.74 &  6.88 &   0.30 &  0.03 &  0.29 \\ 
   PG 1101+249 & $-$8.63 & $-$0.09 &   +0.36 &  $-$33 & +256  & $-$56 &   +30  & +256 & +2209 &  13.02 &  8.47 &   1.53 &  0.12 &  0.21 \\ 
   PG 1114+073 & $-$8.58 & $-$0.21 &   +0.39 &      0 & +197  &  $-$4 &  $-$5  & +197 & +1688 &   8.60 &  7.15 &   0.41 &  0.05 &  0.10 \\ 
 PG 1232$-$136 & $-$8.33 & $-$0.33 &   +0.43 &  $-$79 & +135  &   +38 &   +73  & +138 & +1152 &   9.17 &  3.72 &   0.82 &  0.09 &  0.42 \\ 
   PG 1233+427 & $-$8.56 &   +0.06 &   +0.31 &    +17 & +227  &   +74 & $-$15  & +227 & +1942 &  10.11 &  8.47 &   1.78 &  0.18 &  0.09 \\ 
      Feige 66 & $-$8.50 & $-$0.01 &   +0.18 &    +24 & +217  &    +9 & $-$24  & +217 & +1841 &   9.10 &  7.76 &   0.22 &  0.02 &  0.08 \\ 
   PG 1256+278 & $-$8.48 &   +0.02 &   +0.78 &  $-$71 & +197  &   +75 &   +72  & +196 & +1666 &  10.68 &  6.33 &   2.19 &  0.21 &  0.26 \\ 
 PG 1343$-$101 & $-$8.13 & $-$0.27 &   +0.55 &  $-$32 & +149  &   +58 &   +27  & +150 & +1217 &   8.29 &  4.42 &   1.31 &  0.16 &  0.30 \\ 
     HD 127493 & $-$8.42 & $-$0.05 &   +0.07 &    +10 & +212  &   +15 & $-$11  & +212 & +1780 &   8.53 &  7.71 &   0.20 &  0.02 &  0.05 \\ 
   PG 1432+004 & $-$8.05 & $-$0.08 &   +0.61 &    +34 & +143  & $-$20 & $-$36  & +143 & +1151 &   8.55 &  3.68 &   0.69 &  0.08 &  0.36 \\ 
   PG 1433+239 & $-$8.34 &   +0.10 &   +0.43 &    +14 & +189  & $-$44 & $-$12  & +189 & +1574 &   8.41 &  6.47 &   0.90 &  0.11 &  0.13 \\ 
   PG 1452+198 & $-$8.14 &   +0.16 &   +0.71 &    +71 & +169  &   +51 & $-$67  & +171 & +1388 &   9.09 &  5.13 &   1.48 &  0.16 &  0.28 \\ 
   PG 1519+640 & $-$8.58 &   +0.44 &   +0.47 &  $-$54 & +332  & $-$93 &   +71  & +329 & +2826 &  29.98 &  8.36 &   4.89 &  0.20 &  0.56 \\ 
   PG 1619+522 & $-$8.41 &   +0.55 &   +0.53 &  $-$31 & +199  & $-$24 &   +44  & +197 & +1658 &   9.40 &  7.05 &   0.63 &  0.08 &  0.17 \\ 
     HD 149382 & $-$8.43 &   +0.01 &   +0.04 &    +13 & +233  &   +11 & $-$13  & +233 & +1966 &   9.64 &  8.33 &   0.13 &  0.01 &  0.07 \\ 
   PG 1647+252 & $-$8.10 &   +0.40 &   +0.43 &  $-$11 & +265  &   +43 &   +24  & +264 & +2142 &  12.83 &  8.01 &   1.14 &  0.09 &  0.23 \\ 
   PG 1708+602 & $-$8.48 &   +1.45 &   +1.05 & $-$108 & +164  &   +95 &  +134  & +143 & +1232 &  12.05 &  3.78 &   3.75 &  0.31 &  0.52 \\ 
   PG 1710+490 & $-$8.35 &   +0.56 &   +0.42 &    +27 & +208  & $-$54 & $-$13  & +210 & +1755 &   8.94 &  8.05 &   1.00 &  0.13 &  0.05 \\ 
   PG 1716+426 & $-$8.12 &   +0.91 &   +0.68 &   +128 & +211  & $-$44 &  +105  & +230 & +1877 &  12.73 &  6.27 &   1.38 &  0.10 &  0.35 \\ 
   PG 1722+286 & $-$8.04 &   +0.59 &   +0.44 &  $-$44 & +222  &   +14 &   +61  & +218 & +1760 &   9.77 &  6.20 &   0.50 &  0.06 &  0.19 \\ 
   PG 1725+252 & $-$8.11 &   +0.43 &   +0.32 &  $-$51 & +180  &   +33 &   +61  & +177 & +1435 &  10.08 &  8.01 &   1.64 &  0.06 &  0.27 \\ 
   PG 1738+505 & $-$8.32 &   +0.80 &   +0.51 &  $-$27 & +241  &   +53 &   +50  & +238 & +1987 &  11.56 &  7.68 &   1.37 &  0.12 &  0.20 \\ 
   UV0 1758+36 & $-$8.42 &   +0.16 &   +0.08 &     +5 & +224  &   +33 &  $-$1  & +224 & +1889 &   8.89 &  8.42 &   0.46 &  0.05 &  0.03 \\ 
     HD 171858 & $-$8.34 &   +0.03 & $-$0.02 &    +86 & +227  &     0 & $-$85  & +227 & +1894 &  11.74 &  6.69 &   0.03 &  0.0025 &  0.27 \\ 
HE 2135$-$3749 & $-$8.02 &   +0.05 & $-$0.54 & $-$109 & +244  &  +113 &  +110  & +243 & +1951 &  16.11 &  6.83 &   4.90 &  0.32 &  0.40 \\ 
     HD 205805 & $-$8.37 & $-$0.02 & $-$0.15 &  $-$82 & +225  &    +3 &   +82  & +225 & +1880 &  11.48 &  6.72 &   0.19 &  0.02 &  0.20 \\ 
HE 2154$-$4143 & $-$7.58 & $-$0.01 & $-$1.17 &  $-$32 & +270  &  $-$7 &   +32  & +270 & +2047 &  13.04 &  7.50 &   1.78 &  0.16 &  0.27 \\ 
HE 2155$-$1724 & $-$7.84 &   +0.51 & $-$0.96 &  $-$54 & +203  & $-$18 &   +67  & +199 & +1563 &   9.42 &  5.90 &   1.11 &  0.12 &  0.23 \\ 
HE 2156$-$1732 & $-$7.78 &   +0.56 & $-$1.05 &  $-$51 & +136  &   +15 &   +61  & +132 & +1028 &   8.33 &  3.41 &   1.18 &  0.14 &  0.42 \\ 
HE 2156$-$3927 & $-$7.65 &   +0.04 & $-$1.11 & $-$153 & +370  &  +141 &  +155  & +369 & +2822 &  61.63 &  7.10 &  21.26 &  0.37 &  0.79 \\ 
HE 2201$-$2136 & $-$7.52 &   +0.63 & $-$1.47 &  $-$56 & +183  & $-$30 &   +71  & +178 & +1341 &   8.93 &  5.50 &   1.72 &  0.20 &  0.24 \\ 
   PG 2204+035 & $-$8.11 &   +0.82 & $-$0.76 &    +20 & +261  & $-$85 &    +6  & +262 & +2132 &  13.74 &  8.15 &   3.01 &  0.22 &  0.26 \\ 
HE 2205$-$1952 & $-$8.00 &   +0.36 & $-$0.80 &  $-$31 & +172  &   +19 &   +38  & +170 & +1363 &   8.33 &  5.12 &   0.92 &  0.11 &  0.24 \\ 
HE 2213$-$4158 & $-$6.25 & $-$0.08 & $-$3.25 & $-$104 & +406  & $-$71 &   +99  & +407 & +2544 &  62.40 &  7.02 &  25.40 &  0.44 &  0.80 \\ 
   PG 2218+020 & $-$8.16 &   +0.76 & $-$0.80 &    +43 & +201  & $-$41 & $-$24  & +205 & +1676 &   8.56 &  7.27 &   1.21 &  0.14 &  0.08 \\ 
HE 2222$-$3738 & $-$7.76 &   +0.07 & $-$1.17 & $-$227 & +147  &   +12 &  +228  & +145 & +1123 &  16.53 &  2.89 &   2.55 &  0.16 &  0.70 \\ 
   PG 2226+094 & $-$8.27 &   +0.84 & $-$0.72 &  $-$64 & +184  &     0 &   +83  & +177 & +1468 &  12.30 &  7.17 &   1.53 &  0.08 &  0.32 \\ 
   PG 2259+134 & $-$8.43 &   +1.03 & $-$0.91 &    +39 &  +20  & $-$44 & $-$14  & +212 & +1803 &   8.79 &  8.22 &   1.38 &  0.16 &  0.03 \\ 
     Feige 109 & $-$8.46 &   +0.75 & $-$0.84 &   $-$7 &  +24  &   +63 &   +29  & +240 & +2033 &  12.50 &  8.38 &   1.85 &  0.19 &  0.16 \\ 
     Feige 108 & $-$8.45 &   +0.22 & $-$0.33 &    +43 & +238  & $-$33 & $-$37  & +239 & +2022 &  10.95 &  8.05 &   0.75 &  0.07 &  0.15 \\ 
 PG 2337$-$070 & $-$8.53 &   +0.48 & $-$0.60 &   +138 & +154  & $-$28 &  +129  & +162 & +1381 &  11.60 &  4.32 &   1.12 &  0.10 &  0.46 \\ 
HE 2337$-$2944 & $-$8.28 &   +0.08 & $-$0.84 &  $-$63 & +241  &   +16 &   +65  & +240 & +1986 &  11.96 &  7.42 &   1.22 &  0.10 &  0.23 \\ 
HE 2340$-$2806 & $-$8.16 &   +0.17 & $-$1.40 &  $-$38 & +298  &   +24 &   +44  & +298 & +2428 &  18.82 &  8.09 &   3.09 &  0.17 &  0.40 \\ 
        SB 815 & $-$8.43 &   +0.00 & $-$0.24 &    +44 & +238  &  $-$7 & $-$43  & +238 & +2009 &  10.89 &  7.86 &   0.30 &  0.03 &  0.16 \\ 
HE 2343$-$2944 & $-$8.20 &   +0.11 & $-$1.25 &    +20 & +136  &    +2 & $-$18  & +137 & +1120 &   8.31 &  3.88 &   1.26 &  0.15 &  0.36 \\ 
HE 2349$-$3135 & $-$8.02 &   +0.10 & $-$2.02 &   +117 & +254  &$-$151 &  +114  & +256 & +2050 &  22.52 &  7.68 &  11.84 &  0.61 &  0.49 \\ 
   PG 2349+002 & $-$8.52 &   +0.42 & $-$0.70 &    +76 & +163  &   +61 & $-$68  & +167 & +1422 &   9.69 &  5.16 &   1.62 &  0.17 &  0.31 \\ 
        SB 884 & $-$8.42 &   +0.02 & $-$0.36 &  $-$28 & +191  &  $-$1 &   +28  & +191 & +1605 &   8.69 &  6.34 &   0.37 &  0.04 &  0.16 \\ 
HE 2355$-$3221 & $-$8.14 &   +0.04 & $-$1.65 &    +78 & +126  & $-$43 & $-$77  & +126 & +1026 &   9.00 &  3.52 &   2.33 &  0.26 &  0.44 \\ 
   PG 2358+107 & $-$8.62 &   +0.52 & $-$0.63 &    +49 & +193  &  $-$9 & $-$37  & +196 & +1691 &   9.18 &  6.73 &   0.69 &  0.08 &  0.15 \\ 
HE 2359$-$2844 & $-$8.23 &   +0.12 & $-$1.50 &  $-$23 & +122  &  +125 &   +24  & +122 & +1002 &   8.35 &  4.79 &   4.92 &  0.70 &  0.27 \\ 